\documentclass[a4paper,11pt]{article}
\pdfoutput=1 
\usepackage{jcappub}
\usepackage{amsmath}
\usepackage{graphicx}
\usepackage{latexsym}
\usepackage{xspace}
\usepackage{color}
\usepackage{hyperref} 
\usepackage{bm}
\usepackage{relsize}
\usepackage{tabularx}
\usepackage{multirow}
\usepackage{amssymb}
\usepackage[table]{xcolor}
\usepackage{slashbox}
\usepackage{placeins}
\usepackage{afterpage}

\makeatletter
\gdef\@fpheader{}
\g@addto@macro\bfseries{\boldmath}
\makeatother

\newcommand{\ie}{{i.e.~}}

\newcommand{\eg}{{e.g.~}}

\newcommand{\dd}{\mathrm{d}}
\newcommand{\ee}{e}

\newcommand{\sss}[1]{{\scriptscriptstyle{#1}}}

\newcommand{\uPl}{\mathrm{Pl}}

\newcommand{\umin}{\mathrm{min}}
\newcommand{\umax}{\mathrm{max}}
\newcommand{\uend}{\mathrm{end}}

\newcommand{\ureh}{\mathrm{reh}}

\newcommand{\urad}{\mathrm{rad}}

\newcommand{\uc}{\mathrm{c}}

\newcommand{\uS}{\mathrm{S}}

\newcommand{\usssS}{\sss{\uS}}

\newcommand{\usssPl}{\sss{\uPl}}

\newcommand{\nS}{n_\usssS}

\newcommand{\calP}{\mathcal{P}}

\newcommand{\Mp}{M_\usssPl}

\newcommand{\efolds}{$e$-folds}
\newcommand{\efold}{$e$-fold}

\newcommand{\beq}{\begin{equation}}
\newcommand{\eeq}{\end{equation}}
\newcommand{\bea}{\begin{eqnarray}}
\newcommand{\eea}{\end{eqnarray}}

\newlength{\wsingfig}
\newlength{\wdblefig}
\newlength{\wquadfig}
\newlength{\wtriplefig}
\newcommand{\Eq}[1]{Eq.~(\ref{#1})}
\newcommand{\Eqs}[1]{Eqs.~(\ref{#1})}
\newcommand{\Fig}[1]{Fig.~{\ref{#1}}}

\newcommand{\Ref}[1]{Ref.~{\cite{#1}}}
\newcommand{\Refs}[1]{Refs.~{\cite{#1}}}
\newcommand{\Sec}[1]{Sec.~\ref{#1}}
\newcommand{\Secs}[1]{Secs.~\ref{#1}}

\subheader{}

\title{Geometrical destabilization, premature end of inflation and Bayesian model selection}

\author[a,b]{S\'ebastien Renaux-Petel,}
\affiliation[a]{Institut d'Astrophysique de Paris, UMR 7095 du CNRS, Sorbonne Universit\'es et UPMC Univ. Paris 6, 98 bis bd Arago, 75014 Paris, France}
\affiliation[b]{Sorbonne Universit\'es, Institut Lagrange de Paris, 98 bis bd Arago, 75014 Paris, France}
\emailAdd{renaux@iap.fr}

\author[c]{Krzysztof Turzy\'nski,}
\affiliation[c]{Institute of Theoretical Physics, Faculty of Physics, University of Warsaw, Pasteura 5, 02-093 Warsaw, Poland}
\emailAdd{krzysztof.turzynski@fuw.edu.pl}

\author[d,e]{Vincent Vennin}
\affiliation[d]{Laboratoire Astroparticule et Cosmologie, Universit\'e Denis Diderot Paris 7, 75013 Paris, France}
\emailAdd{vincent.vennin@port.ac.uk}
\affiliation[e]{Institute of Cosmology \& Gravitation, University of Portsmouth, Dennis Sciama Building, Burnaby Road, Portsmouth, PO13FX, United Kingdom}

\date{today}

\begin{document}
\sloppy

\abstract{
By means of Bayesian techniques, we study how a premature ending of inflation, motivated by geometrical destabilization, affects the observational evidences of typical inflationary models. Large-field models are worsened, and inflection point potentials are drastically improved for a specific range of the field-space curvature characterizing the geometrical destabilization. For other models we observe shifts in the preferred values of the model parameters. For quartic hilltop models for instance, contrary to the standard case, we find preference for theoretically natural sub-Planckian hill widths. Eventually, the Bayesian ranking of models becomes substantially reordered with a premature end of inflation. Such a phenomenon also modifies the constraints on the reheating expansion history, which has to be properly accounted for since it determines the position of the observational window with respect to the end of inflation. Our results demonstrate how the interpretation of cosmological data in terms of fundamental physics is considerably modified in the presence of premature end of inflation mechanisms.}


\arxivnumber{1706.01835}

\maketitle

\flushbottom

\section{Introduction}
\label{sec:intro}
Inflation~\cite{Starobinsky:1980te, Sato:1980yn, Guth:1980zm, Linde:1981mu, Albrecht:1982wi, Linde:1983gd} describes a phase of accelerated expansion in the early Universe, during which vacuum quantum fluctuations of the gravitational and matter fields were amplified to cosmological perturbations~\cite{Starobinsky:1979ty, Mukhanov:1981xt, Hawking:1982cz,  Starobinsky:1982ee, Guth:1982ec, Bardeen:1983qw}. These primordial fluctuations later seeded the cosmic microwave background (CMB) anisotropies and the large-scale structure of our Universe.

At present, the full set of observations can be accounted for in a minimal setup, where inflation is driven by a single scalar inflaton field with canonical kinetic term, minimally coupled to gravity, and evolving in a flat potential in the slow-roll regime~\cite{Ade:2015lrj,Ade:2015ava,Martin:2013tda, Martin:2013nzq,Renaux-Petel:2015bja}. In most of these models, the potential becomes steeper as inflation proceeds and inflation eventually stops when the potential is not flat enough to support it. 

However, several mechanisms can lead to a premature end of inflation (PEI). In hybrid inflation~\cite{Linde:1991km, Linde:1993cn, Copeland:1994vg, Lyth:1998xn, Panagiotakopoulos:2000ky, Lazarides:2000ck, Covi:2000gx} for instance, an auxiliary field $\chi$ is coupled to the inflaton field $\phi$ through a term $\propto \phi^2 \chi^2$, so that the effective mass of $\chi$ is $\phi$-dependent. At early time during inflation, $\chi$ has a large mass compared to the Hubble scale and it is a mere spectator field. As $\phi$ rolls down its potential, the mass squared of $\chi$ decreases and eventually becomes negative. This triggers a tachyonic instability that quickly terminates inflation (see however \Refs{Clesse:2010iz, Avgoustidis:2011em, Martin:2011ib} for cases where this ``waterfall'' phase extends over several \efolds).

Recently, the geometrical destabilization (GD) of inflation has offered a new mechanism that can prematurely terminate inflation \cite{Renaux-Petel:2015mga}. The GD is the phenomenon by which the field space curvature can dominate forces originating from the potential and destabilize inflationary trajectories. Similarly to the well-known eta problem (see for instance \Ref{Baumann:2014nda}), it is a manifestation of the sensitivity of inflation to the physics near the Planck scale. It thus represents both a universal challenge and an opportunity to learn about the highest energy scales from cosmological observations. In the following, we make use of a minimal realization of the GD outlined in \Ref{Renaux-Petel:2015mga}. A well-motivated possible outcome of this phenomenon is that it terminates inflation abruptly, much earlier than what slow-roll violation would have yielded. When this premature end of inflation occurs, the location of the observational window along the inflationary effective single-field potential is modified. This means that the part of the potential that the inflaton field is exploring when the modes of astrophysical interest today crossed the Hubble radius changes, and so do the predictions of a given model of inflation.

In the present work, we confront the predictions of the models with a premature end of inflation with data, using Bayesian techniques. In practice, we consider various classes of prototypical models that are differently affected by such a phenomenon, which allow us to discuss most possible effects resulting from a change in the end of inflation time. Since the location of the observational window is also affected by the expansion history of reheating, which determines how far the observational window is from the end of inflation, we carefully incorporate this epoch and describe its degeneracies with the end of inflation location. The paper is organized as follows. In \Sec{sec:method}, we introduce the physical setup and the method used in this work. The geometrical destabilization is presented in \Sec{sec:GD}, the role played by reheating in \Sec{sec:reheating}, the Bayesian model comparison approach in \Sec{sec:Bayesian} and the prototypical models we use in \Sec{sec:models}. In \Sec{sec:results}, we analyze our results, both at the level of posterior distributions on the parameters of the models and regarding their relative Bayesian evidences. Our main conclusions are summarized in \Sec{sec:concl}.
\section{Setup and method}
\label{sec:method}
\subsection{Geometrical destabilization and premature end of inflation}
\label{sec:GD}
The geometrical destabilization of inflation is a generic phenomenon that potentially affects all realistic models embedded in high-energy physics. For simplicity, we only make use in this paper of a minimal realization of it through the following phenomenological two-field Lagrangian:
\bea
\mathcal{L} = -\left(1+2\frac{\chi^2}{M_{\mathcal{R}}^2}\right)\frac{\left(\partial\phi\right)^2}{2}
-V\left(\phi\right)-\frac{\left(\partial\chi\right)^2}{2}
-\frac{m_h^2}{2}\chi^2\, .
\label{minimal}
\eea
Let us first consider the situation in which the interaction $\propto  (\partial \phi)^2\chi^2$ is absent. The Lagrangian (\ref{minimal}) then describes an inflaton field $\phi$, slowly rolling down its potential $V(\phi)$, with an additional scalar field $\chi$ which is heavy during inflation for $m_h^2 \gg H^2$ (where $H\equiv \dot{a}/a$ is the Hubble scale, $a$ is the scale factor and a dot denotes derivation with respect to cosmic time), so that it is anchored at the bottom of the inflationary valley at $\chi=0$. Let us now assess the impact of the kinetic coupling $-(\partial \phi)^2\chi^2/M_{\mathcal{R}}^2$, whose presence is generic from the effective field theory point of view, and where $M_{\mathcal{R}}$ denotes the energy scale associated with new physics above the energy scale of inflation $H$.\footnote{The subscript ${\mathcal{R}}$ comes from the fact that the kinetic coupling induces a curvature of the field space, the Ricci scalar of which is related to $M_{\mathcal{R}}$ through $\mathcal{R} = -4/M_{\mathcal{R}}^2$ when $\chi \simeq 0$.}
Along the inflationary valley, what can appear merely as a small correction to the kinetic term of $\phi$ provides a negative, time-dependent contribution to the effective mass of $\chi$, which reads 
\bea
m_{\mathrm{eff}}^2= m_h^2 - 4 \epsilon(t) H^2(t) \Mp^2/M_{\mathcal{R}}^2,
\label{eq:meff}
\eea
where $\Mp$ is the reduced Planck mass and $\epsilon\equiv \dot{\phi}^2/(2 H^2 \Mp^2)$ is the first slow-roll parameter. If $\epsilon(t) H^2(t)$ increases during inflation, $m_{\mathrm{eff}}^2$ can turn from positive to negative, triggering an instability when $\epsilon$ reaches the critical value
\bea
\label{eq:epsilonc}
\epsilon_{ \uc} =  \frac{1}{4} \left(\frac{m_h}{H_\uc}\right)^2 \left(\frac{M_\mathcal{R}}{\Mp}\right)^2\, .
\eea
If $\epsilon_{ \uc}<1$, this instability occurs before inflation ends by slow-roll violation, and can lead to a premature ending of inflation, see \Refs{Renaux-Petel:2015mga,preparation} for a discussion on the fate of the instability. As argued there, the energy scale $M_\mathcal{R}$ associated to the field-space curvature can a priori lie anywhere between the Hubble scale and the Planck scale, so that $\epsilon_{ \uc}$ can be orders of magnitude smaller than one, and the GD arises in the bulk of the would-be inflationary phase.
With the assumption of the instability prematurely ending inflation, a powerful lower bound on $M_\mathcal{R}$ also arises from the requirement to obtain the correct power spectrum amplitude~\cite{Ade:2015xua} $\calP_\zeta = 2.2\times 10^{-9}$ of curvature fluctuations at the pivot scale $k_* = 0.05\,\mathrm{Mpc}^{-1}$. At leading order in the slow-roll approximation, it is given by $\calP_\zeta = H_*^2/(8 \pi^2 \Mp^2 \epsilon_*)$, where $H_*$ and $\epsilon_*$ are evaluated at the time when $k_*$ crosses the Hubble radius during inflation (all the quantities with a subscript ``*'' are evaluated at that time). At that time, $m_{\mathrm{eff}}^2$ is still positive, which implies that 
\bea
\frac{M_\mathcal{R}}{H_*} > \frac{1}{\sqrt{2 \pi^2 \calP_\zeta}} \frac{H_*}{m_h} \simeq 5000 \frac{H_*}{m_h}.
\label{eq:lowerbound:MR}
\eea
Note that a similar bound on the mass scale characterizing the field space curvature is derived in \Ref{Renaux-Petel:2015mga} beyond the simple setup described by the Lagrangian \Eq{eq:DeltaNstar}, which shows how the GD offers the possibility to constrain interactions at
energies well above $H$, and the internal geometry of high-energy physics theories.

Notice that in the minimal realization used here, $\epsilon(t) H^2(t)$, or equivalently $\dot{\phi}^2$, must increase as time proceeds, which may not be satisfied for some inflationary potentials, 
\eg $V(\phi)\propto\phi^p$ with $p\geq2$. However, in more general setups, $m_h$ may receive corrections of order $H$ (see \eg \Ref{Dine:1995uk}) or
$M_\mathcal{R}$ may depend on $\phi$ and the GD may still provide a viable way to end inflation prematurely. To leave these possibilities open, we will consider a PEI for any choice of potential $V(\phi)$, parameterize it by the value $\epsilon_{\mathrm{c}}$ of the slow-roll parameter at the end of inflation, and use \Eq{eq:epsilonc} to interpret it in the context of the GD for the models that allow it.
\subsection{Reheating}
\label{sec:reheating}
As mentioned in \Sec{sec:intro}, the expansion history realized during reheating determines the amount of expansion realized between the Hubble exit time of the scales probed in the CMB and the end of inflation, hence the location of the observational window along the inflationary potential. More precisely, the number of \efolds~$\Delta N_*=\ln(a_\uend/a_*)$ elapsed between Hubble exit time of the pivot scale $k_*$ and the end of inflation is given by~\cite{Martin:2006rs, Martin:2010kz, Easther:2011yq}
\bea
\Delta N_* = \frac{1-3\bar{w}_\ureh}{12\left(1+\bar{w}_\ureh\right)}\ln\left(\frac{\rho_\ureh}{\rho_\uend}\right)
+\frac{1}{4}\ln\left(\frac{\rho_*}{9\Mp^4}\frac{\rho_*}{\rho_\uend}\right)
-\ln\left(\frac{k_{*}/a_\mathrm{now}}{\tilde{\rho}_{\gamma,\,\mathrm{now}}^{1/4}}\right)\, .
\label{eq:DeltaNstar}
\eea
In this expression, $\bar{w}_\ureh=\int_\ureh w(N)\dd N/N_\ureh$ is the averaged equation-of-state parameter during reheating, $\rho_\ureh$ is the energy density of the Universe at the end of reheating, \ie at the beginning of the radiation-dominated era, $\rho_*$ is the energy density calculated $\Delta N_*$ \efolds~before the end of inflation, $\rho_\uend$ is the one at the end of inflation, $a_\mathrm{now}$ is the present value of the scale factor, and $\tilde{\rho}_{\gamma,\,\mathrm{now}}$ is the energy density of radiation today rescaled by the number of relativistic degrees of freedom. 

Reheating should occur after inflation and before big-bang nucleosynthesis (BBN), such that $\rho_\mathrm{BBN}<\rho_\ureh<\rho_\uend$. Although in certain situations the reheating temperature may be below $1\,\mathrm{MeV}$
\cite{Kawasaki:1999na, Kawasaki:2000en}, in this work we make the conservative choice $\rho^{1/4}_{\mathrm{BBN}} = 10\, \mathrm{MeV}$. Moreover, from the energy positivity conditions in general relativity, and the definition of reheating as the era following the accelerated phase of expansion, one has $-1/3<\bar{w}_\mathrm{reh}<1$. By letting $\rho_\ureh$ and $\bar{w}_\mathrm{reh}$ vary in their respective ranges, from \Eq{eq:DeltaNstar} one obtains an interval of possible values for $\Delta N_*$, hence a certain uncertainty range along the inflationary potential. Notice that this interval depends on the inflationary potential, since in \Eq{eq:DeltaNstar}, the way $\rho_*/\rho_\uend$ is related to $\Delta N_*$ depends on the inflationary history, hence on the potential, and the absolute value of $\rho_*$ is constrained by $\calP_\zeta$, which also involves $\epsilon_*$, that itself depends on the potential as well.
\begin{figure}[t]
\begin{center}
\includegraphics[width=0.5\textwidth]{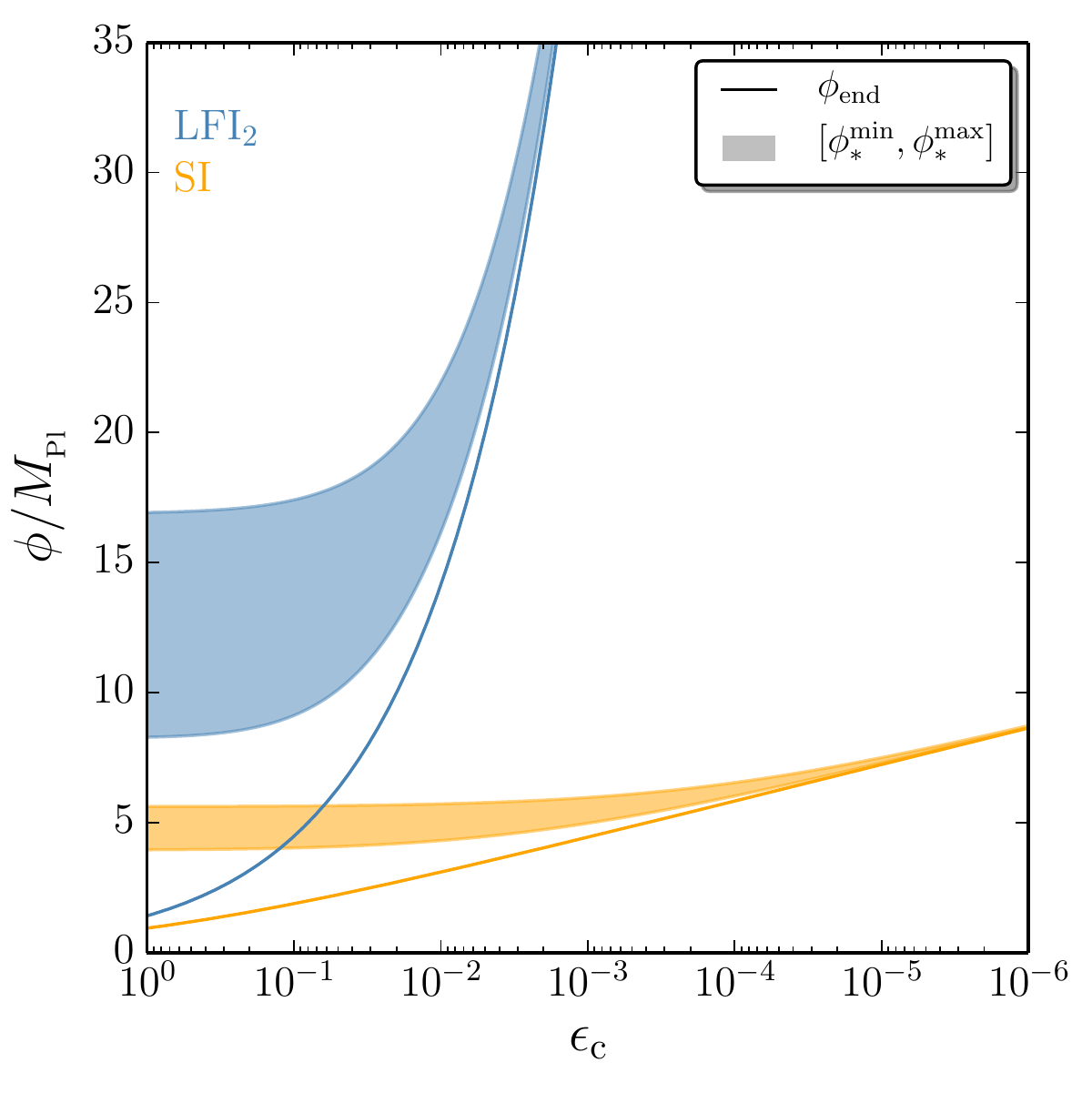}
\caption{Value of the inflaton field at the end of inflation $\phi_\uend$ (solid lines) and uncertainty range corresponding to the pivot scale $[\phi_{*}^{\umin}, \phi_{*}^{\umax}]$ (shaded stripes), as a function of the value $\epsilon_\uc$ taken by $\epsilon$ at the end of inflation, for two of the models studied in this work: $\mathrm{LFI}_2$ (large-field inflation $V\propto \phi^2$, blue) and $\mathrm{SI}$ (the Starobinsky model, orange), see \Sec{sec:models}. For each value of $\epsilon_\uc$, the range of allowed values for $\Delta N_*$ is calculated using \Eq{eq:DeltaNstar}. This is then translated into a range of allowed values for $\phi_*$ displayed with the colored stripes. This figure shows that the location of the observational window depends both on $\epsilon_\uc$, which determines the value of $\phi_\uend$, and on the reheating parameters through $\Delta N_*$ given in \Eq{eq:DeltaNstar}.
\label{fig:phiBound}}
\end{center}
\end{figure}
To illustrate this effect and its degeneracy with a premature end of inflation, in \Fig{fig:phiBound}, we have represented the relevant field values as a function of $\epsilon_\uc$ for two of the models studied in this work. For a given value of $\epsilon_\uc$, which determines the value of $\phi_\uend$, we can infer from \Eq{eq:DeltaNstar} the range of allowed values for $\phi_*$ corresponding to the pivot scale $k_*$ of the CMB; this range is displayed with the colored stripes. The location of that range depends both on the PEI parameter $\epsilon_\uc$ and on the reheating parameters through $\Delta N_*$ given in \Eq{eq:DeltaNstar}. Naturally, the more premature the end of inflation is, the less important the effects of the uncertainties on the reheating expansion history are. The range $[\phi (\epsilon=1) , \phi (\epsilon=\epsilon_\uc)]$  (below the line) is entirely removed  from the inflationary dynamics, while the range  $[\phi(\epsilon=\epsilon_\uc), \phi_{*}^{\umin}]$ (between the line and the stripe) is dynamically accessible but corresponds to smaller scales than the pivot scale.  The range $[\phi_{*}^{\umin}, \phi_{*}^{\umax}]$ is probed at the pivot scale, where the uncertainty comes from our incomplete knowledge of reheating.
\subsection{Bayesian model comparison}
\label{sec:Bayesian}
In order to discuss the impact of a PEI on identifying the favored models of inflation, we use the Bayesian inference techniques~\cite{Cox:1946,Jeffreys:1961, deFinetti:1974, Trotta:2005ar, Trotta:2008qt} developed in \Refs{Martin:2013tda, Martin:2013nzq, Martin:2014nya, Vennin:2015eaa, Martin:2016oyk}. In this framework, for each model $\mathcal{M}_i$ (labeled by $i$), the posterior probability $p$ of its parameters $\theta_{ij}$ (labeled by $j$) is expressed as
\bea
\label{eq:posterior:def}
p\left(\theta_{ij}\vert\mathcal{D},\mathcal{M}_i\right)=\frac{\mathcal{L}
\left(\mathcal{D}\vert\theta_{ij},\mathcal{M}_i\right)
\pi\left(\theta_{ij}\vert \mathcal{M}_i\right)}{\mathcal{E}\left(\mathcal{D}\vert\mathcal{M}_i \right) } \, .
\eea
In this expression, $\mathcal{L}(\mathcal{D}\vert\theta_{ij},\mathcal{M}_i)$ is the likelihood and represents the probability of observing the data $\mathcal{D}$ assuming the model $\mathcal{M}_i$ is true and $\theta_{ij}$ are the actual values of its parameters, $\pi (\theta_{ij}\vert \mathcal{M}_i )$ is the prior distribution on the parameters $\theta_{ij}$, and $\mathcal{E}\left(\mathcal{D}\vert\mathcal{M}_i \right)$ is a normalization constant called the Bayesian evidence and defined as
\bea
\label{eq:evidence:def}
\mathcal{E}\left(\mathcal{D}\vert\mathcal{M}_i \right) 
= \int\dd\theta_{ij}\mathcal{L}
\left(\mathcal{D}\vert\theta_{ij},\mathcal{M}_i\right)
\pi\left(\theta_{ij}\vert \mathcal{M}_i\right)\, .
\eea
The Bayesian evidence allows one to calculate the posterior probability of a model itself, $p(\mathcal{M}_i\vert\mathcal{D})\propto \mathcal{E}(\mathcal{D}\vert\mathcal{M}_i)\pi(\mathcal{M}_i)$, where $\pi(\mathcal{M}_i)$ is the prior assigned to the model. The posterior odds between two models $\mathcal{M}_i$ and $\mathcal{M}_j$ can then be written as
\bea
\frac{p\left(\mathcal{M}_i\vert\mathcal{D}\right)}
{p\left(\mathcal{M}_j\vert\mathcal{D}\right)}
=\frac{\mathcal{E}\left(\mathcal{D}\vert\mathcal{M}_i\right)}
{\mathcal{E}\left(\mathcal{D}\vert\mathcal{M}_j\right)}
\frac{\pi\left(\mathcal{M}_i\right)}{\pi\left(\mathcal{M}_j\right)}\equiv
B_{ij}\frac{\pi\left(\mathcal{M}_i\right)}{\pi\left(\mathcal{M}_j\right)}\,, 
\eea
where we have defined the Bayes factor $B_{ij}$ by $B_{ij}=\mathcal{E}\left(\mathcal{D}\vert\mathcal{M}_i\right) /\mathcal{E}\left(\mathcal{D}\vert\mathcal{M}_j\right)$. Under the principle of indifference, one can assume non-committal model priors, $\pi(\mathcal{M}_i)=\pi\left(\mathcal{M}_j\right)$, in which case the Bayes factor becomes identical to the posterior odds. With this assumption, a Bayes factor larger (smaller) than one means a preference for the model $\mathcal{M}_i$ over the model $\mathcal{M}_j$ (a preference for $\mathcal{M}_j$ over $\mathcal{M}_i$). In practice, the ``Jeffreys' scale'' gives an empirical prescription for translating the values of the Bayes factor into strengths of belief. When $\ln(B_{ij})>5$, $\mathcal{M}_j$ is said to be ``strongly disfavored'' with respect to $\mathcal{M}_i$, ``moderately disfavored'' if $2.5<\ln(B_{ij})<5$, ``weakly disfavored'' if $1<\ln(B_{ij})<2.5$, and the situation is said to be ``inconclusive'' if $\ln(B_{ij})<1$. Bayesian analysis allows us to identify the models that achieve the best compromise between quality of the fit and lack of fine tuning.

In practice, the data $\mathcal{D}$ used in this work is the Planck 2015 $TT$ data combined with the high-$\ell$ $C_\ell^{TE}+C_\ell^{EE}$ likelihood and the low-$\ell$ temperature plus polarization likelihood (PlanckTT,TE,EE+lowTEB in the notations of \Ref{Aghanim:2015xee}, see table~1 there), together with the BICEP2-Keck/Planck likelihood described in \Ref{Ade:2015tva}, and the effective likelihood via slow-roll reparameterization of \Ref{Ringeval:2013lea} is employed. The predictions of the models are computed making use of the publicly available \texttt{ASPIC} library~\cite{aspic}, which has been extended to incorporate the GD.

An important aspect of Bayesian analysis is the role played by priors. For the GD sector, assuming that the scale of new physics $M_{\mathcal{R}}$ is sub-Planckian, but leaving its order of magnitude undetermined, we adopt a Jeffreys prior (\ie logarithmically flat) $-25<\log_{10}(M_{\mathcal{R}}/\Mp)<0$. The bound~(\ref{eq:lowerbound:MR}) seems to impose an additional hard prior condition, but the normalization of the power spectrum automatically takes care of it.
In practice, when $M_{\mathcal{R}}/\Mp>2 H_\uc/m_h$, \Eq{eq:epsilonc} yields a value for $\epsilon_\uc$ that is larger than one, which means that the GD does not occur and inflation ends by slow-roll violation in a standard way.
One may be concerned that the lower bound $10^{-25}\Mp$ corresponds to energy scales small enough to be probed in particle physics experiments. In a cosmological context however, the Friedmann equation $H^2=\rho/(3\Mp^2)$ involves the Planck mass and $\rho_{\mathrm{BBN}}^{1/4} = 10\, \mathrm{MeV}$ yields a lower bound on $H_*$ (hence on $M_{\mathcal{R}}$) of the order $H_{\mathrm{BBN}} \sim 10^{-42} \Mp$, so $10^{-25}\Mp$ is in fact more than conservative in this sense. 
In our parameterization, $m_h$ only appears divided by $H_\uc$ in the expression~(\ref{eq:epsilonc}) of $\epsilon_\uc$. In the following, we use the value $m_h/H_\uc=10$ but one should note that assuming $m_h/H_\uc$ to take another fixed value, or allowing it to vary with a logarithmically flat prior, would not modify the shape of the effective prior on $\epsilon_\uc$ induced by \Eq{eq:epsilonc} that is logarithmically flat. It would only change its lower bound but that would not affect our main conclusions. Let us also stress that using an effective logarithmically flat prior on $\epsilon_\uc$ allows us to scan premature end of inflation in general, beyond the phenomenon of the GD.

For the reheating sector, $\rho_{\ureh}$ and $\bar{w}_\ureh$ only appear in \Eq{eq:DeltaNstar} through the combination given in the first term of the right-hand side denoted $\ln R_\urad = (1-3 \bar{w}_\ureh)/(12+12\bar{w}_\ureh) \ln(\rho_\ureh/\rho_\uend)$. As explained in \Sec{sec:reheating}, $\rho_\ureh$ can vary between $\rho_{\mathrm{BBN}}$ and $\rho_\uend$, and $\bar{w}_\ureh$ can vary between $-1/3$ and $1$, so that $\ln(\rho_{\mathrm{BBN}}/\rho_\uend)/4<\ln R_\urad<\ln(\rho_\uend/\rho_{\mathrm{BBN}})/12$. Since the order of magnitude of $R_\urad$ is unknown between these two bounds, they define a logarithmically flat prior on $R_\urad$.

Thus far we have specified all priors except for the parameters specifying the model of inflation one considers. We now turn to the presentation of the prototypical models used in our analysis and the treatment of their free parameters.
\subsection{Prototypical models}
\label{sec:models}
\begin{figure}[!ht]
\begin{center}
\includegraphics[width=0.328\textwidth]{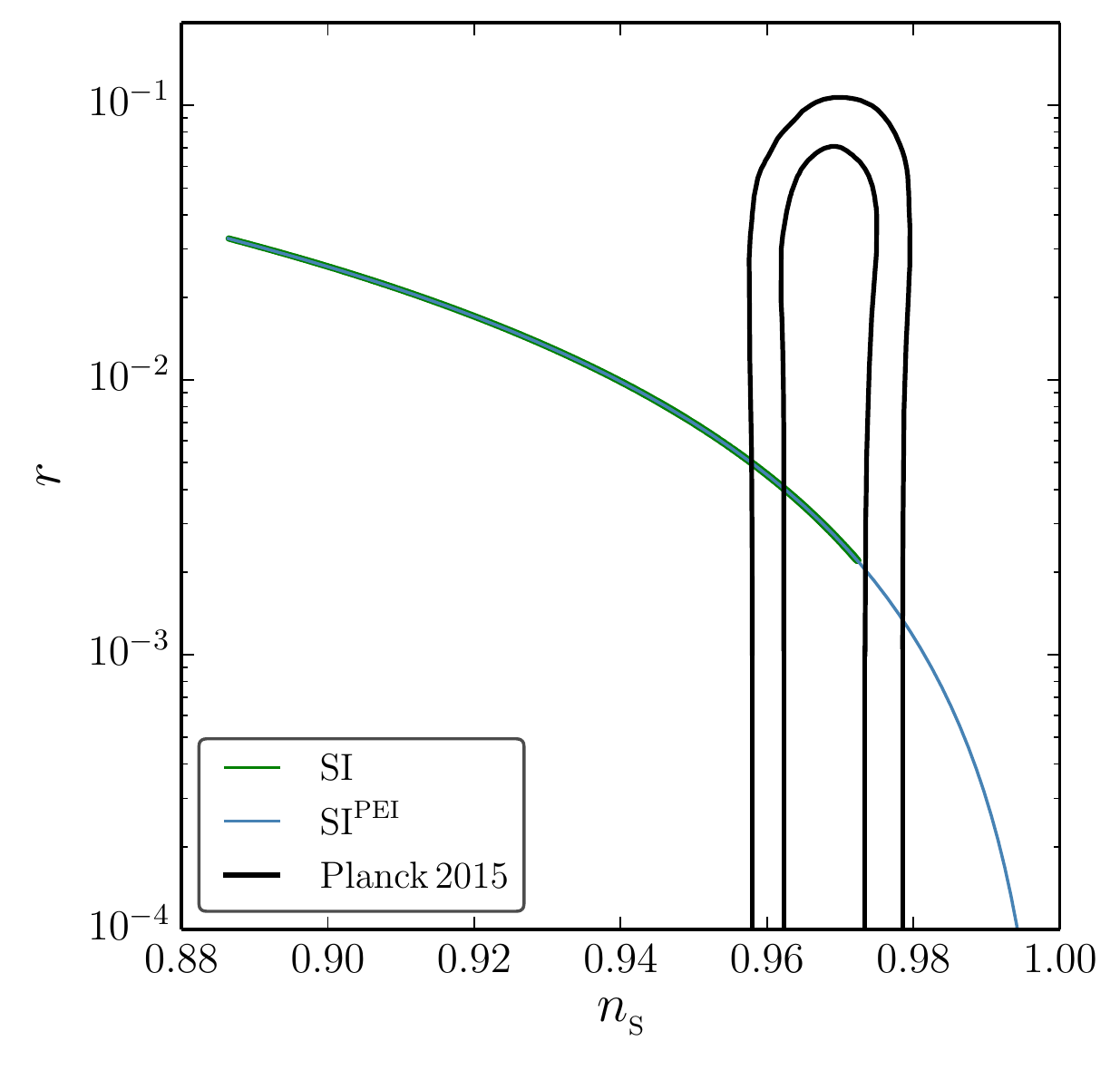}
\includegraphics[width=0.328\textwidth]{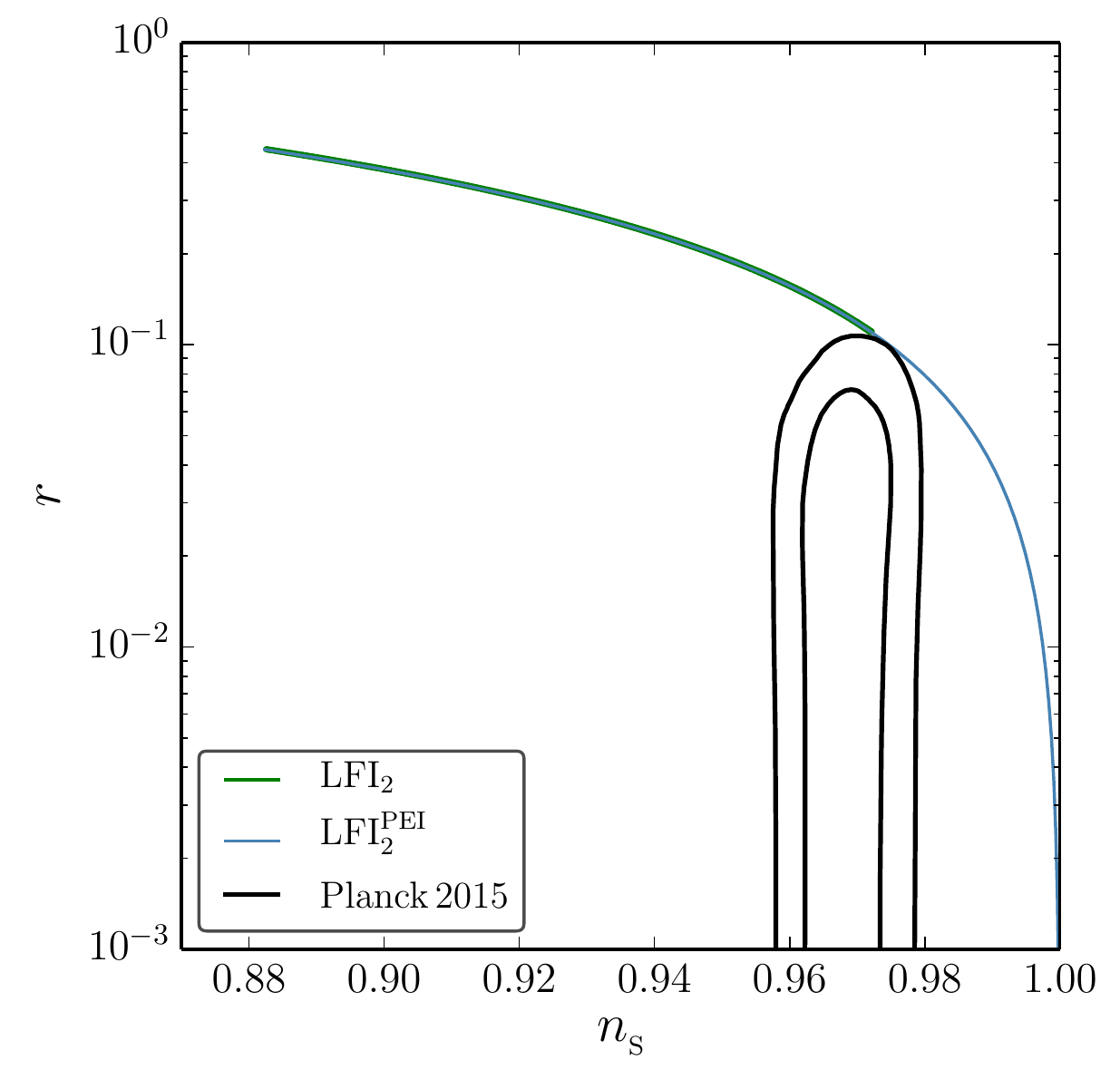}
\includegraphics[width=0.328\textwidth]{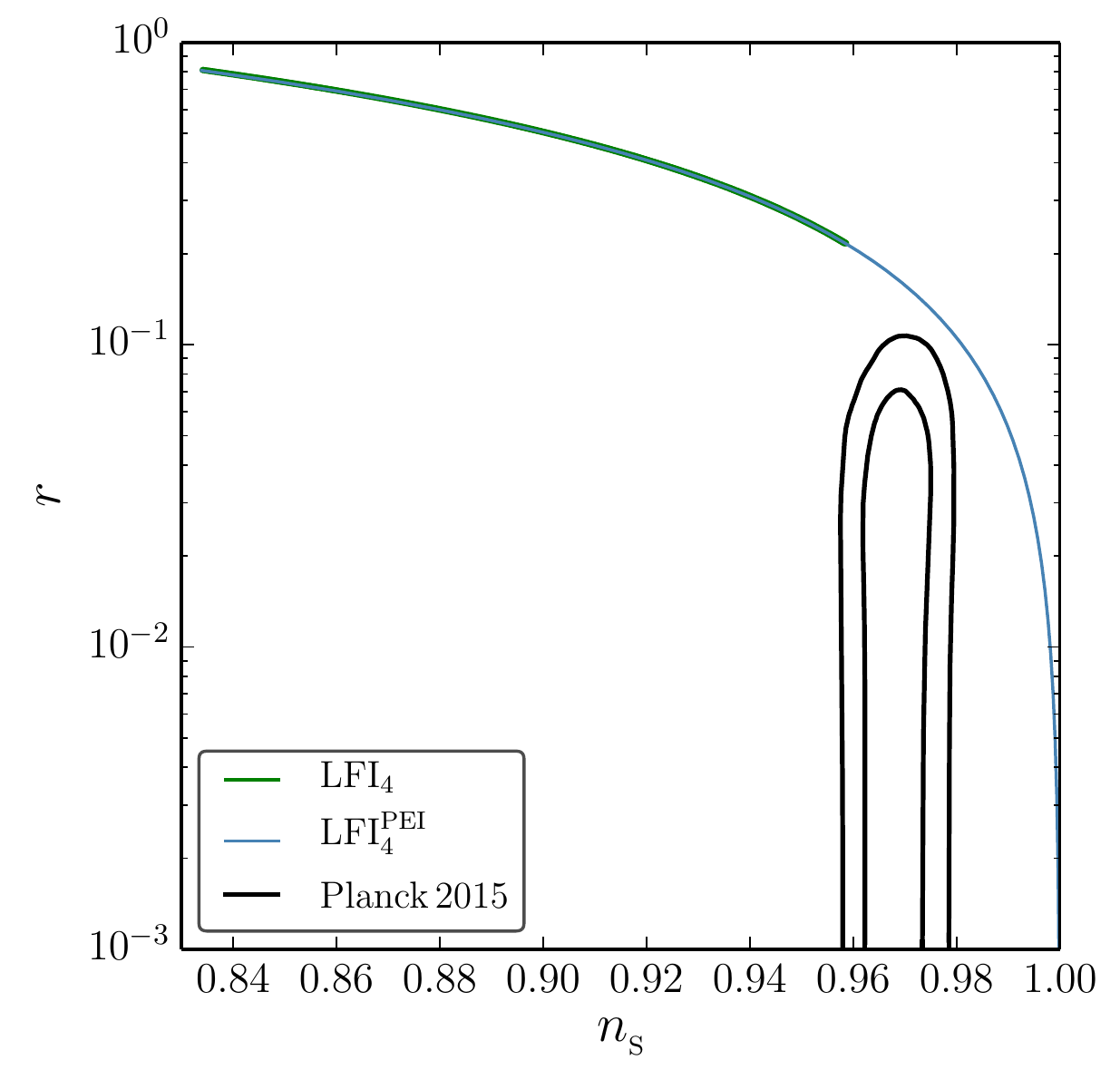}
\includegraphics[width=0.328\textwidth]{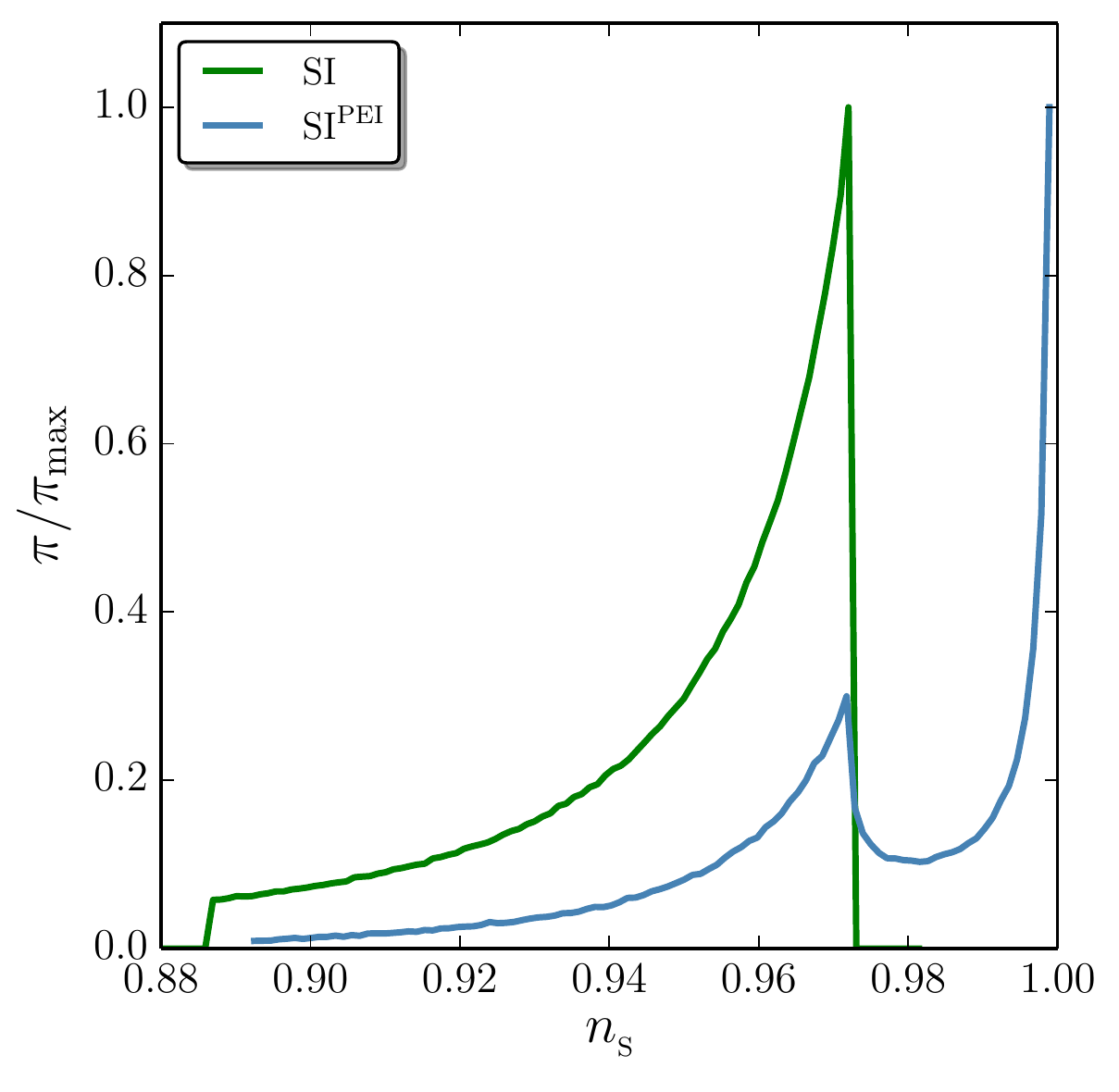}
\includegraphics[width=0.328\textwidth]{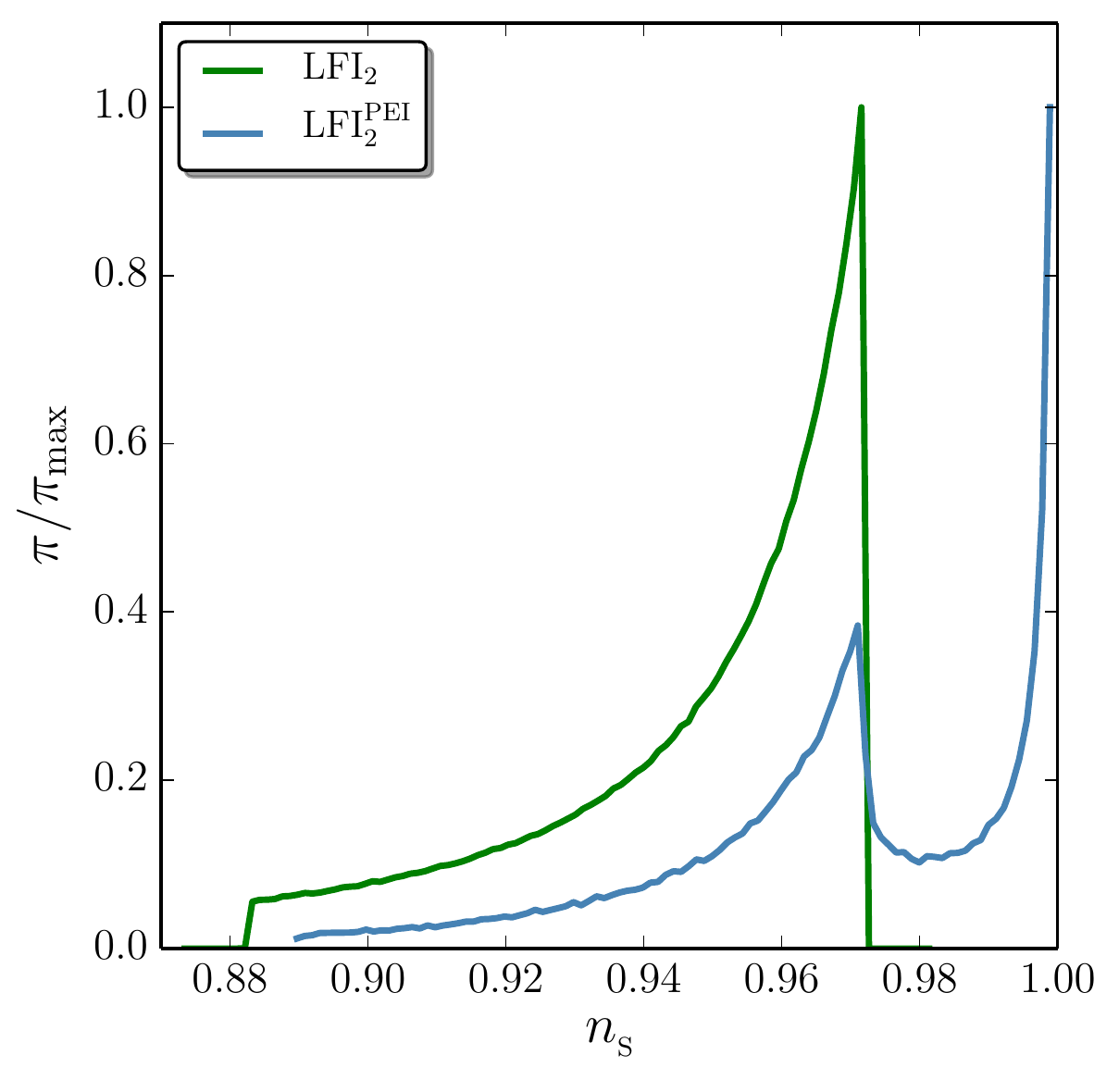}
\includegraphics[width=0.328\textwidth]{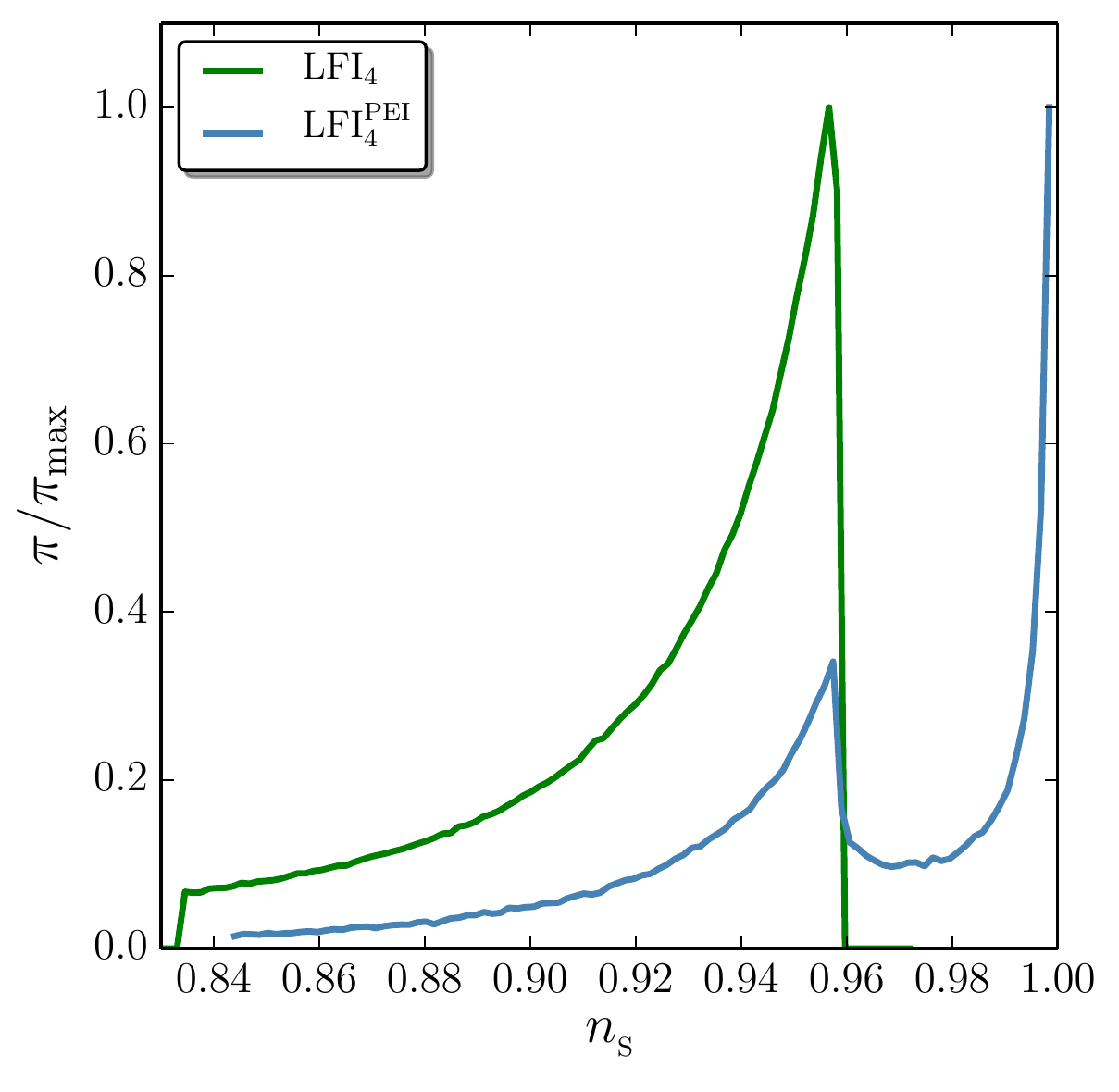}
\includegraphics[width=0.328\textwidth]{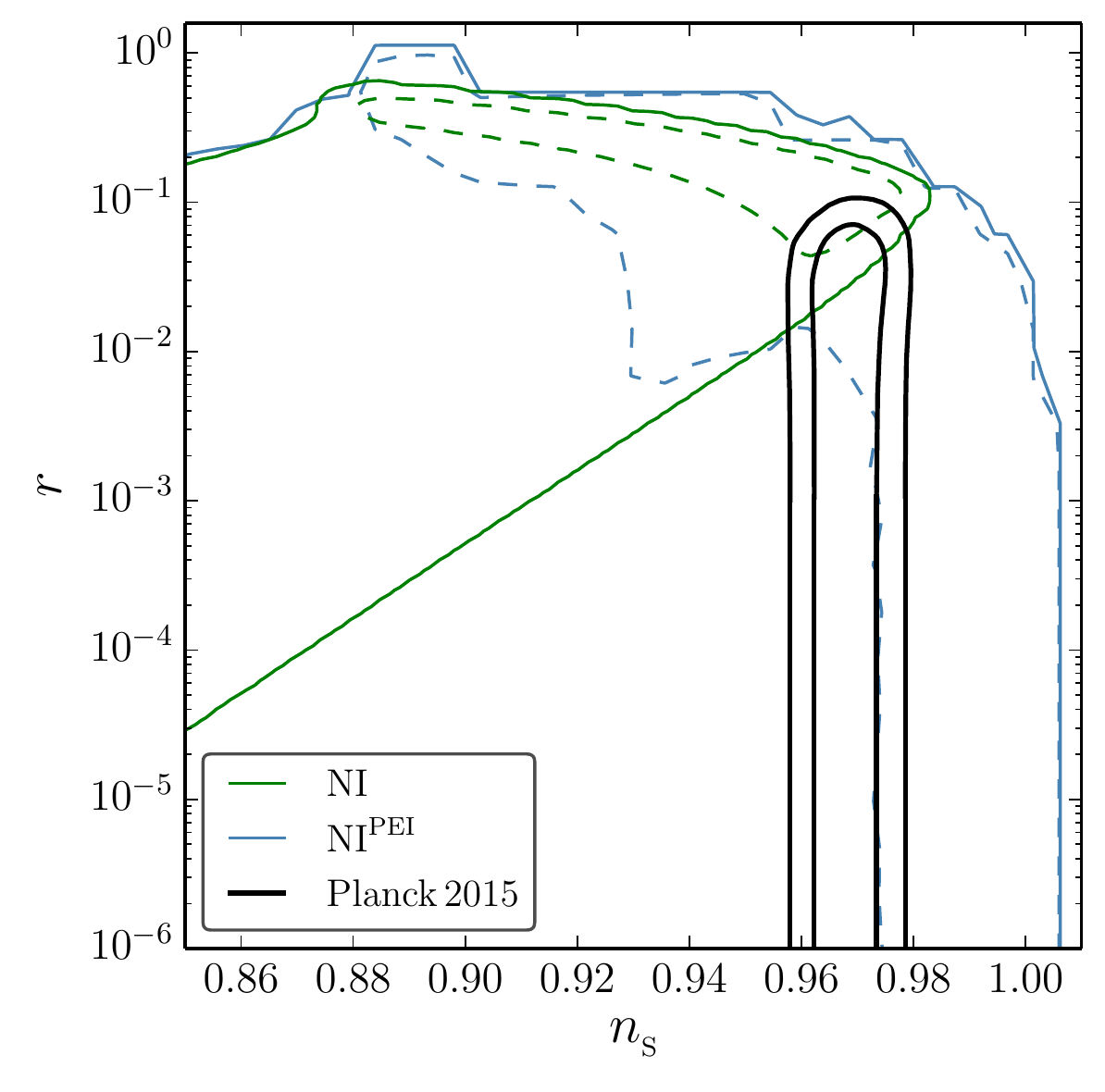}
\includegraphics[width=0.328\textwidth]{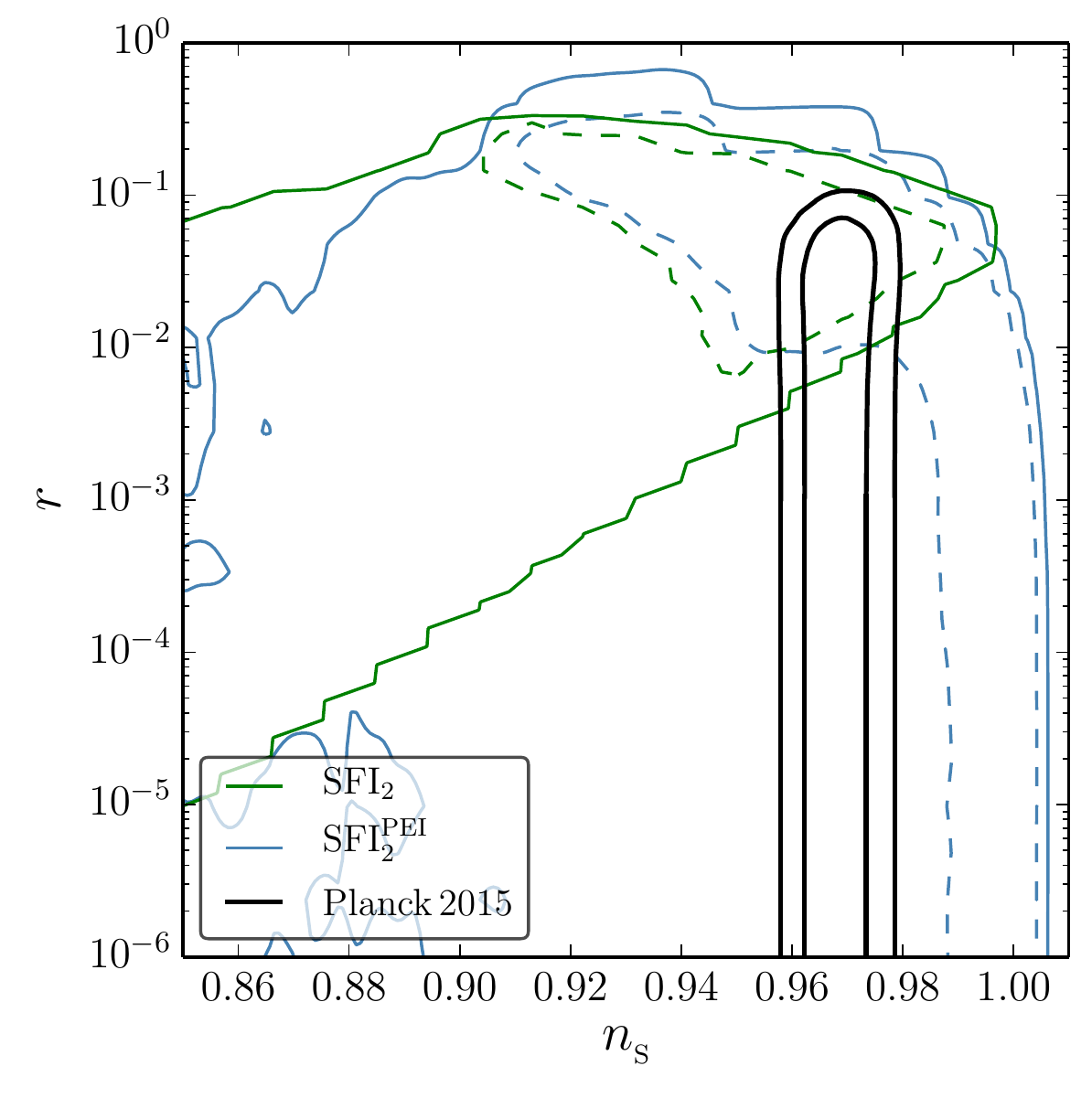}
\includegraphics[width=0.328\textwidth]{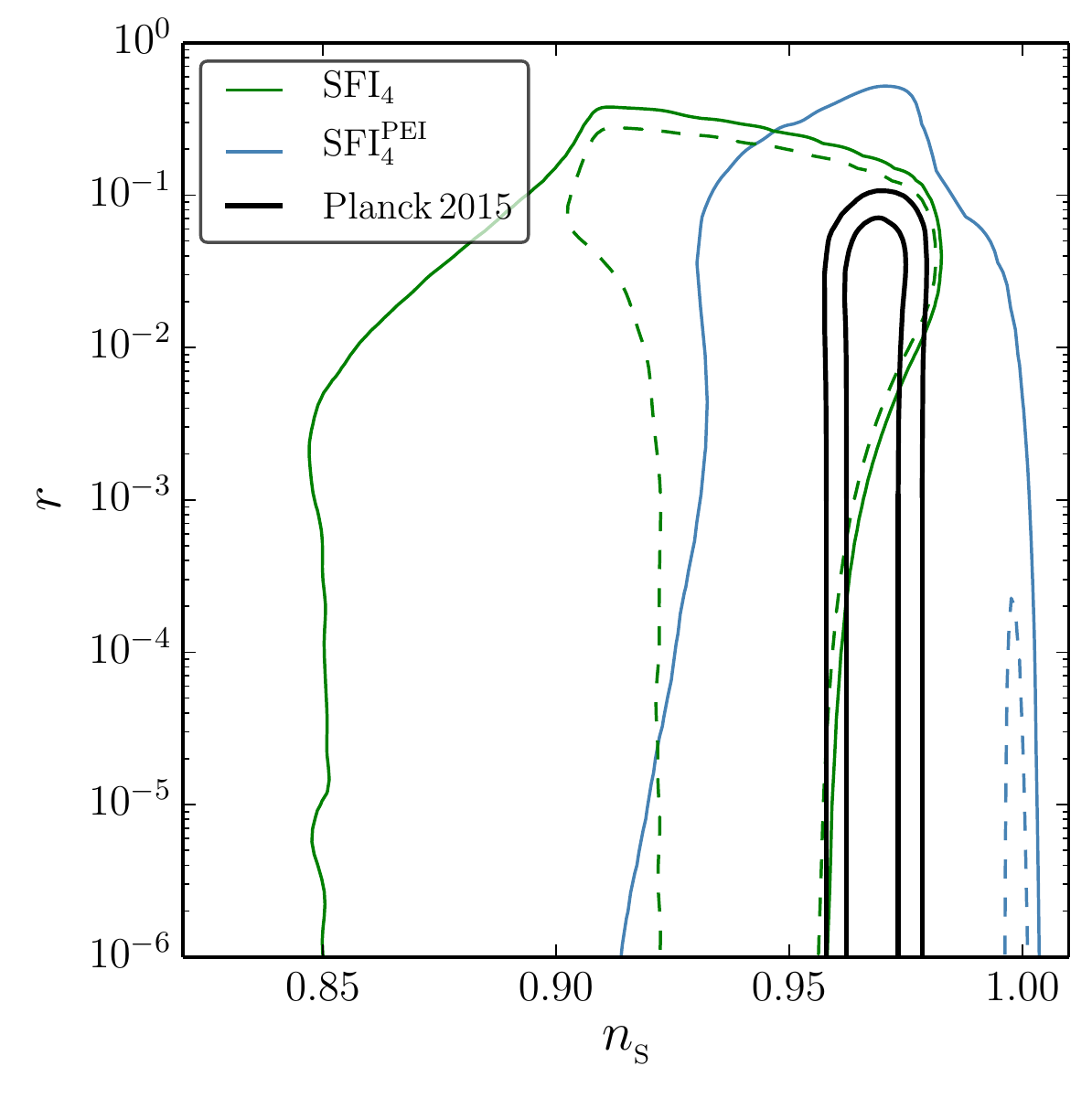}
\includegraphics[width=0.328\textwidth]{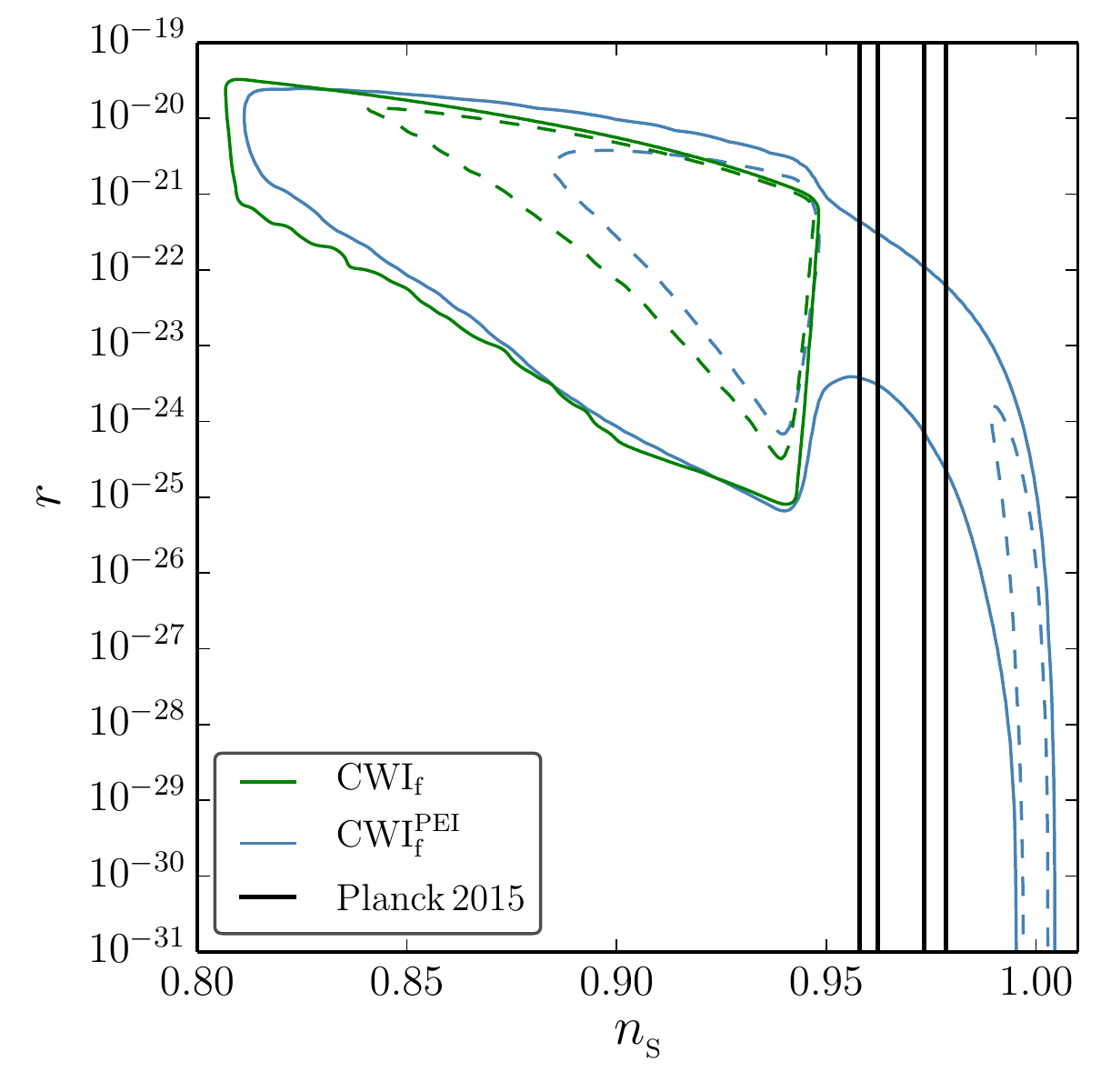}
\includegraphics[width=0.328\textwidth]{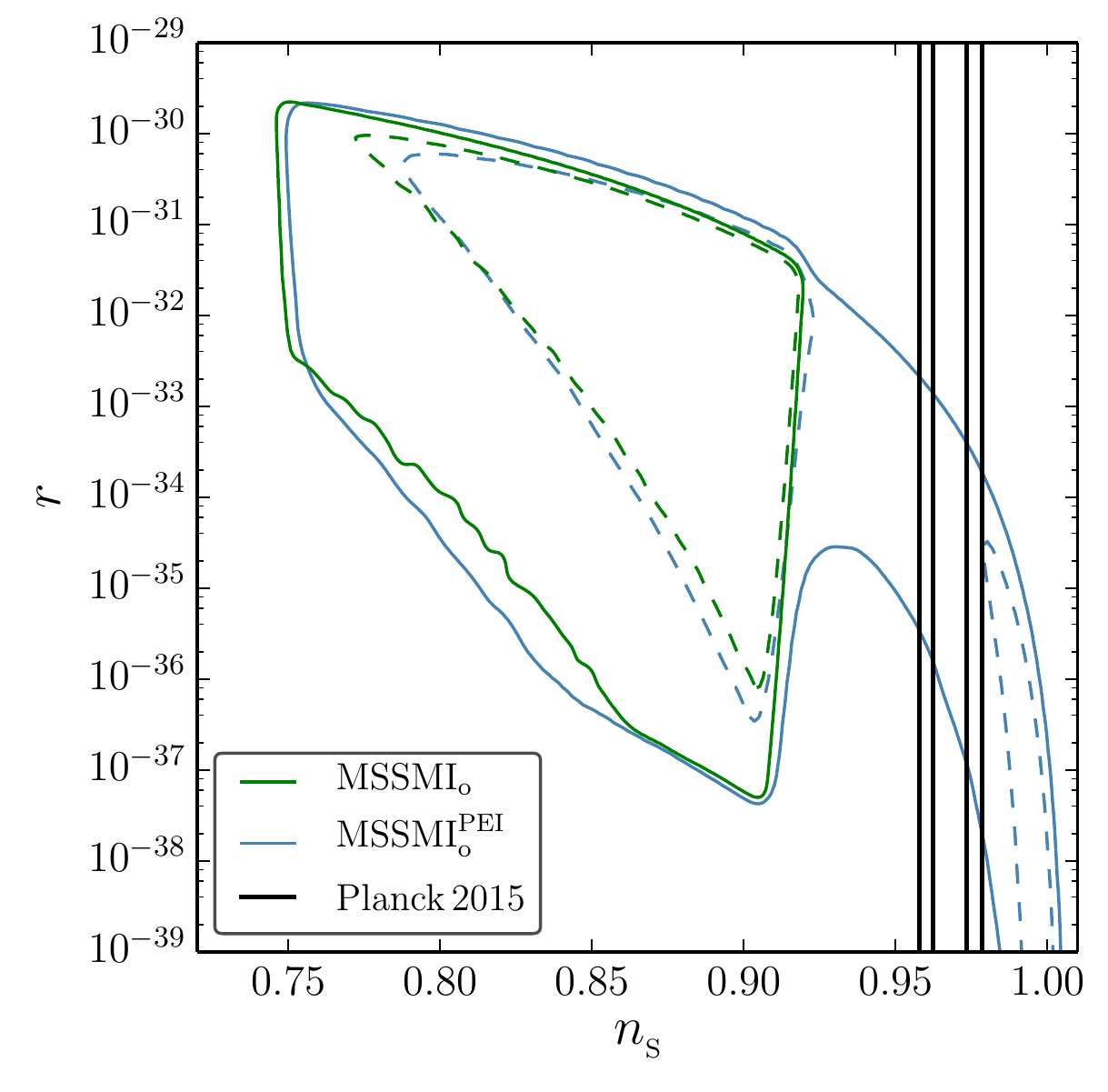}
\includegraphics[width=0.328\textwidth]{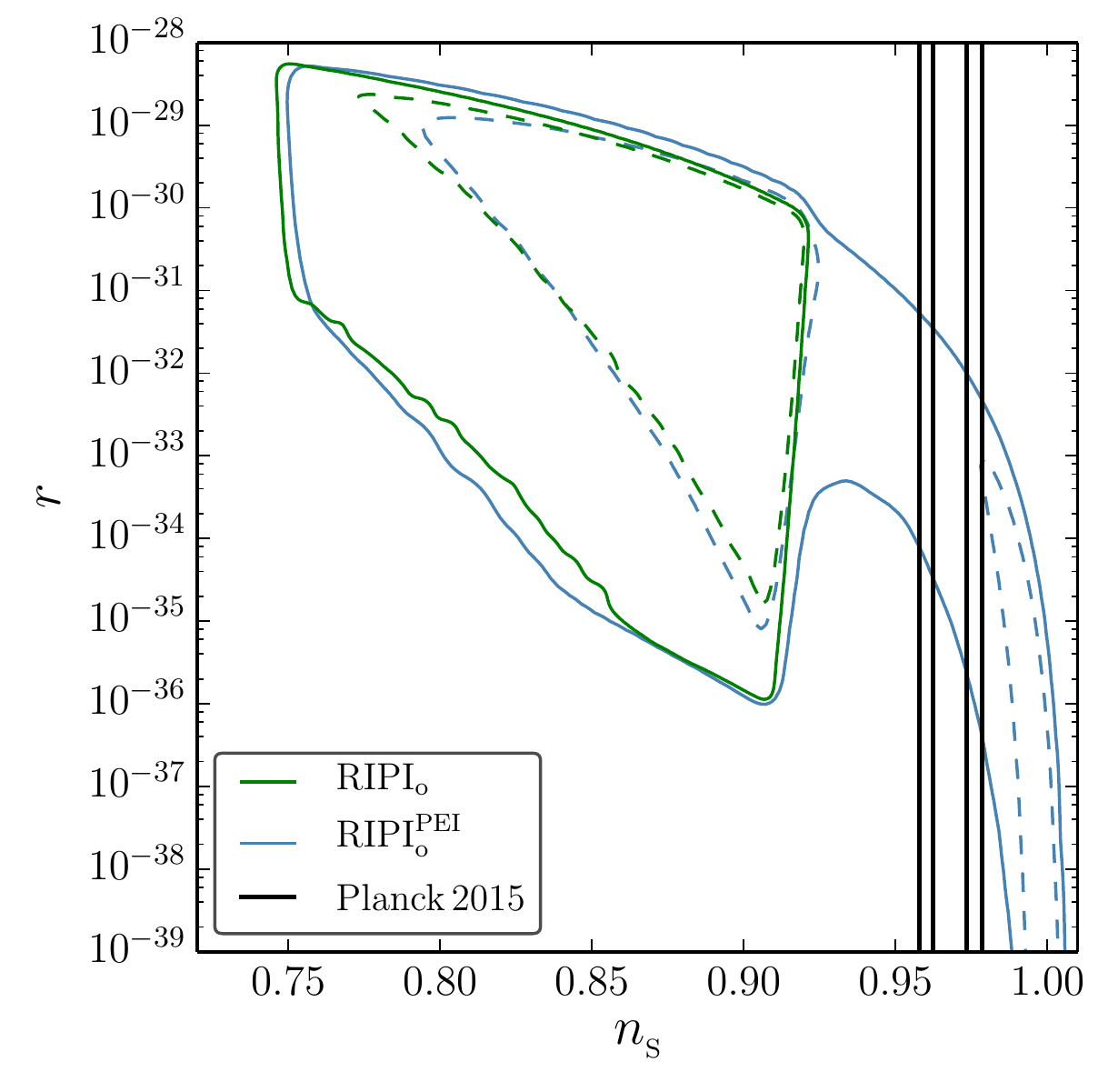}
\caption{Induced priors on $\nS$ and $r$ for the models considered in this work. The solid lines are the two-sigma contours while the dashed lines are the one-sigma contours, the green lines are obtained without PEI (premature end of inflation) and the blue lines with PEI (hence the superscript $\mathrm{PEI}$). The black contours are the one- and two-sigma Planck 2015 constraints. For $\mathrm{SI}$, $\mathrm{LFI}_2$ and $\mathrm{LFI}_4$, there is a one-to-one relationship between $\nS$ and $r$, so the lines simply correspond to all possible predictions without encoding information about their probability densities. The information about the density can be recovered from the priors on $\nS$ alone displayed in the second row of panels for these models.
\label{fig:nsr:prior}}
\end{center}
\end{figure}

Instead of scanning all the $\sim 200$ single-field models reported in \Ref{Martin:2013tda} and implemented in the \texttt{ASPIC} library, our strategy is to identify a few classes of models that behave differently under premature termination of inflation, and to study one or a few prototypical examples in each class.  In order to discuss the predictions of these models and how they compare with the data, before a proper Bayesian analysis is performed in \Sec{sec:results}, in \Fig{fig:nsr:prior} we show the induced priors of these models on $\nS$ and $r$, where $\nS$ is the spectral index of the curvature fluctuation power spectrum $\calP_\zeta$, and $r$ is the tensor-to-scalar ratio, both calculated at the pivot scale $k_*$. In the slow-roll approximation, these are related to the slow-roll parameters through $\nS=1-2\epsilon_1 - \epsilon_2 - 2\epsilon_1^2 - (2C+3)\epsilon_1 \epsilon_2 -C \epsilon_2 \epsilon_3$ and $r= 16\epsilon_1 + 16C \epsilon_1 \epsilon_2$,  where the parameter $C \simeq -0.73$ is a numerical constant. These expressions are valid at second order in slow roll, and the slow-roll parameters $\epsilon_1=\epsilon$, $\epsilon_2$ and $\epsilon_3$ can be calculated from the inflationary potential according to $\epsilon_1 = \Mp^2 (V^\prime/V)^2/2$, $\epsilon_2 = 2 \Mp^2 [(V^\prime/V)^2 -V^{\prime\prime}/V]$ and $\epsilon_3= 2\Mp^4 [V^{\prime\prime\prime} V^\prime/V^2 - 3 V^{\prime\prime} {V^\prime}^2/V^3+ 2 (V^\prime/V)^4] /\epsilon_2$, where these expressions must be evaluated at $\phi=\phi_*$. In practice, the induced priors are calculated using a fiducial constant likelihood in the \texttt{ASPIC} pipeline (one can check that plugging $\mathcal{L}=\mathrm{constant}$ in \Eqs{eq:posterior:def} and~(\ref{eq:evidence:def}) gives $p=\pi$ so that the posteriors extracted in this way indeed correspond to the induced priors). 
\afterpage{\FloatBarrier}
\subsubsection{Plateau models}
\label{sec:model:plateau}
A first class of models is made of plateau potentials that provide a good fit to the data in the standard setup where inflation ends by slow-roll violation. A typical example is the Starobinsky potential (SI)~\cite{Starobinsky:1980te, Bezrukov:2007ep}
\bea
V\left(\phi\right) = M^4\left(1-\ee^{-\sqrt{\frac{2}{3}}\frac{\phi}{\Mp}}\right)^2\, ,
\eea
where, hereafter, the overall mass scale $M^4$ is set to reproduce the correct normalization of $\calP_\zeta$. This potential does not have any free parameter so there is no prior to specify in the inflationary sector. The predictions of this model are displayed in the $(\nS,r)$ plane in the top left panel of \Fig{fig:nsr:prior}. Without a PEI, they fall into the current data sweet spot, with values of $\nS$ that can sometimes be too small but this corresponds to somewhat extreme reheating equation of state parameters $\bar{w}_\ureh$. This can be checked on the prior induced on $\nS$ alone displayed in the left panel of the second row in \Fig{fig:nsr:prior} and which clearly peaks around $\nS\simeq 0.97$. When the PEI is added, the predictions extend to larger values of $\nS$ and smaller values of $r$ that are disfavored by the data.  The prior on $\nS$ is bimodal, with a first peak at the standard predictions and a second one around $\nS\simeq 1$, corresponding to very small values of $\epsilon_\uc$. The relative weights of these two modes depend on the exact lower bound on $\epsilon_\uc$ but one can expect the PEI to decrease the Bayesian evidence of this model in general.
\subsubsection{Large-field models}
\label{sec:model:largefield}
A second class of models is made of large-field potentials that are currently disfavored since they predict values of $r$ that are too large. A typical example is the monomial potential
\bea
V\left(\phi\right) = M^4 \left(\frac{\phi}{\Mp}\right)^p\, ,
\eea
where in this work we consider $p=2$ ($\mathrm{LFI}_2$) and $p=4$ ($\mathrm{LFI}_4$). Their predictions are displayed in the middle and right top panels of \Fig{fig:nsr:prior}, where one can check that in the absence of a premature termination of inflation, the value for $r$ is indeed too large. When a PEI is allowed, $r$ is made smaller, but this is at the expense of making $\nS$ too large, so one can expect these models to remain disfavored even when PEI is added.
\subsubsection{Hilltop models with a non-vanishing mass at the top}
\label{sec:models:hilltopmass}
A third class of models consists in hilltop potentials with $V^{\prime\prime}(\phi=0)\neq 0$. The first example of this class we consider is natural inflation ($\mathrm{NI}$)~\cite{Freese:1990rb, Adams:1992bn}
\bea
\label{eq:pot:ni}
V= M^4 \left[ 1+\cos\left(\frac{\phi}{f}\right)\right]\, .
\eea
From a theoretical perspective, the parameter $f$ is naturally sub-Planckian, but the model provides a good fit to the data only for $f \gtrsim \Mp$, which has motivated various mechanisms proposed in the literature to enlarge the value of $f$ (see \Ref{Baumann:2014nda} for a review). Here we leave the order of magnitude of $f/\Mp$ unspecified, and we work with a logarithmically flat prior $-2<\log_{10}(f/\Mp)<2$. Another example is small-field inflation 2 ($\mathrm{SFI}_2$)
\bea
\label{eq:pot:sfi2}
V\left(\phi\right) = M^4 \left[ 1 - \left(\frac{\phi}{\mu}\right)^2\right]\, ,
\eea
which also has a free parameter $\mu$. If one Taylor expands the potential~(\ref{eq:pot:ni}) of NI and compares it with the one~(\ref{eq:pot:sfi2}) of $\mathrm{SFI}_2$, one can identify $\mu=2f$, which is why we adopt a logarithmically flat prior $-2+\log_{10}(2)<\log_{10}(\mu/\Mp)<2+\log_{10}(2)$ to allow for a fair comparison between these two models. Their predictions are displayed in the left and middle panels of the third row in \Fig{fig:nsr:prior}. In the absence of a PEI, both models are brought to a good agreement with the data when $f/\Mp$ or $\mu/\Mp$ is large. When $f/\Mp$ or $\mu/\Mp$ decreases, $r$ decreases but so does $\nS$ that quickly takes values that are too low. This is because as one approaches the top of the hill, the derivative of the potential becomes very small and so does $r$, but the curvature of the potential saturates to a finite non-vanishing value, which yields a deviation from $\nS$ to $1$ that increases when $f$ or $\mu$ decreases. When a PEI is allowed however, the opposite behavior is observed, since smaller values of $r$ correspond to larger values of $\nS$, that interpolate between the ones favored by the data and $1$, which is excluded. One may therefore expect the Bayesian evidence of these models not to change dramatically by allowing a PEI.
\subsubsection{Hilltop models with a vanishing mass at the top}
\label{sec:models:hilltop:massless}
The behavior of hilltop models is different if $V^{\prime\prime}(\phi=0) = 0$ and this constitutes our fourth class of models. A first example is small-field inflation 4 ($\mathrm{SFI}_4$)
\bea
\label{eq:pot:sfi4}
V\left(\phi\right) = M^4 \left[ 1 - \left(\frac{\phi}{\mu}\right)^4\right]\, ,
\eea
where by consistency with $\mathrm{SFI}_2$, we use the logarithmically flat prior $-2+\log_{10}(2)<\log_{10}(\mu/\Mp)<2+2\log_{10}(2)$. The predictions of $\mathrm{SFI}_4$ are displayed in the right panel of the third row of \Fig{fig:nsr:prior}. They are similar to the ones of $\mathrm{SFI}_2$ except that when $r$ decreases, $\nS$ remains not too far from the observational constraints. When the PEI is allowed, $\nS$ is shifted towards larger values and intersects the ones preferred by the data, even if it shows preference for slightly too large values.  

Another example is the Colemann-Weinberg potential ($\mathrm{CWI}_\mathrm{f}$)~\cite{Coleman:1973jx}
\bea
\label{eq:pot:cwi}
V\left(\phi\right) = M^4 \left[1+\alpha \left(\frac{\phi}{Q}\right)^4\ln\left(\frac{\phi}{Q}\right)\right]\, ,
\eea
where $\alpha=4e$ is a fixed constant set for the potential to vanish at its minimum. In the original version of the scenario, $Q$ is fixed by the GUT scale, $Q \sim 10^{14} - 10^{15}\, \mathrm{GeV}$. It is therefore natural to choose a flat prior on $Q$ (we denote this version of the scenario by $\mathrm{CWI}_\mathrm{f}$, other versions are also considered in \Ref{Martin:2013nzq}), $5\times 10^{-5} < Q/\Mp < 5\times 10^{-4}$. The predictions of $\mathrm{CWI}_\mathrm{f}$ are shown in the bottom left panel of \Fig{fig:nsr:prior}, where one can see that the values predicted for $\nS$ are always too small without a PEI. When a PEI is allowed, $\nS$ is shifted towards larger values as in $\mathrm{SFI}_4$, while $r$ takes smaller values. The one-sigma contours (blue dashed lines) reveal that the distribution is bimodal and is peaked both at the predictions obtained without PEI and at values of $\nS$ close to one, both peaks being observationally disfavored. 
\subsubsection{Inflection point models}
\label{sec:inflection}
The fifth and last class of models we consider is made of potentials with a flat inflection point, such as MSSM inflation ($\mathrm{MSSMI}_\mathrm{o}$)~\cite{Allahverdi:2006iq, Lyth:2006ec}
\bea
\label{eq:pot:mssmi}
V(\phi) = M^4 \left[\left(\frac{\phi}{\phi_0}\right)^2-\frac{2}{3}\left(\frac{\phi}{\phi_0}\right)^6+\frac{1}{5}\left(\frac{\phi}{\phi_0}\right)^{10}\right]\, .
\eea
The free parameter $\phi_0$ can be expressed as $\phi_0^8 = \Mp^6 m_\phi^2/(10 \lambda_6^2)$, where $\lambda_6 $ is a coupling constant that is taken to be of order one, while $m_\phi$ is a soft supersymmetry breaking mass and, thus, is chosen to be around $\simeq 1\, \mathrm{TeV}$. One then obtains $\phi_0\simeq 10^{14}\, \mathrm{GeV}$ and in the original form of this scenario (denoted $\mathrm{MSSMI}_\mathrm{o}$, other versions are also considered in \Ref{Martin:2013nzq}), it is therefore natural to take a flat prior $2\times 10^{-5} < \phi_0/\Mp < 2 \times 10^{-4}$. Another example is the renormalizable inflection point inflation ($\mathrm{RIPI}_\mathrm{o}$) potential~\cite{Allahverdi:2006cx, Allahverdi:2007wt}
\bea
\label{eq:pot:ripi}
V(\phi) = M^4 \left[\left(\frac{\phi}{\phi_0}\right)^2-\frac{4}{3}\left(\frac{\phi}{\phi_0}\right)^3+\frac{1}{2}\left(\frac{\phi}{\phi_0}\right)^4\right]\, ,
\eea
with $\phi_0=\sqrt{3}m_\phi/h$, where $h\simeq 10^{-12}$ is a dimensionless coupling constant and $m_\phi$ is a soft breaking mass of order $100\, \mathrm{GeV} - 10\, \mathrm{TeV}$. One then has $\phi_0\simeq 10^{14}\, \mathrm{GeV}$ as in $\mathrm{MSSMI}_\mathrm{o}$, and the same flat prior $2\times 10^{-5} < \phi_0/\Mp < 2 \times 10^{-4}$ can be used. The predictions of these models are displayed in the middle and right bottom panels of \Fig{fig:nsr:prior}, where the situation is in fact similar to $\mathrm{CWI}_\mathrm{f}$ but even more drastic since the version of the models without PEI predicts values of $\nS$ that are even more disfavored. Otherwise, the same remarks apply here.\\

Before including the observational data in the Bayesian analysis, a final remark is in order. One may be concerned that allowing a PEI brings regions of the potential into the observational window that are so flat that they may be dominated by quantum diffusion effects~\cite{Starobinsky:1982ee, Starobinsky:1986fx}, which would question the consistency of our classical slow-roll approach. In our treatment of the inflection point models presented in \Sec{sec:inflection} for instance, even when $\epsilon_\uc$ takes arbitrarily small values, the observational window is always located below the inflection point, since classically, it takes an infinite number of \efolds~to cross the inflection point. However, quantum diffusion allows the field to cross the inflection point in a finite amount of time, so that one may wonder whether the PEI should extend the observational window to regions located above the inflection point when these stochastic corrections are taken into account. Stochastic effects dominate the field dynamics when the mean quantum kick over one \efold, $H/(2\pi)$, exceeds the classical drift $V^{\prime}/(3H^2)$. Making use of the formula given above \Eq{eq:lowerbound:MR} for $\calP_\zeta$, and recalling that $\epsilon\simeq \Mp^2 (V^\prime/V)^2/2$ in single-field slow-roll inflation, one can see that this happens when $\calP_\zeta>1$. Since the mass scale $M^4$ in the above potentials is set precisely to satisfy the power spectrum normalization condition $\calP_\zeta(k_*)\simeq 2.2\times 10^{-9} \ll 1$, and since the models introduced above are all such that $\calP_\zeta$ decreases as inflation proceeds (they all feature red spectral indices $\nS<1$), one is guaranteed that $\calP_\zeta \ll1$ between $\phi_*$ and $\phi_\uend$, where the field therefore behaves classically, rendering our analysis consistent. In the case of inflection point models, this does not preclude other (larger) viable values of $M^4$ from existing such that the power spectrum would be correctly normalized above the inflection point and that stochastic effects would allow the field to cross the inflection point in the correct finite amount of time, but these solutions are not included in our analysis.
\section{Results}
\label{sec:results}
\begin{figure}[t]
\begin{center}
\includegraphics[width=0.99\textwidth]{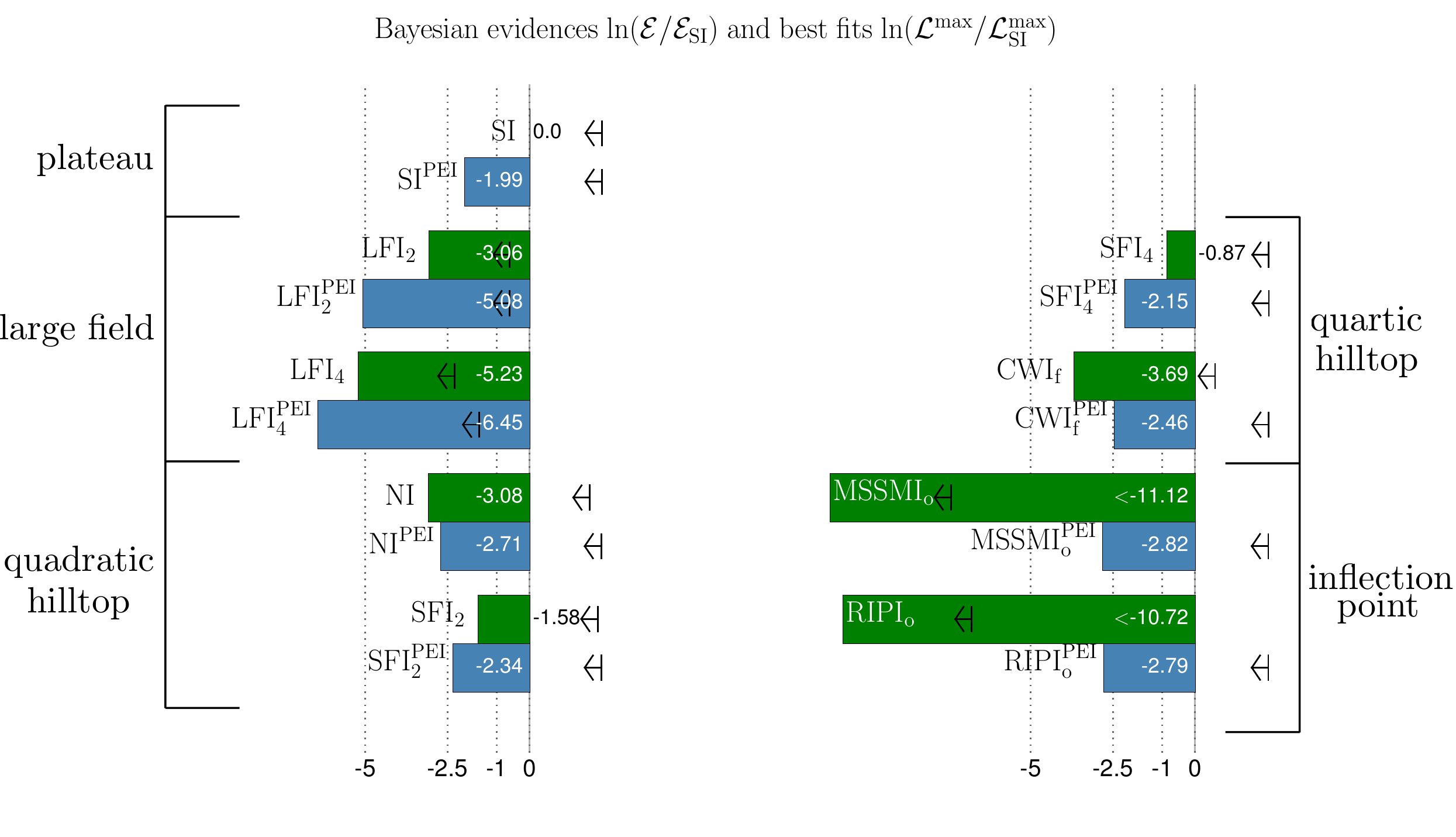}
\caption{Bayesian evidences and best fits for the models considered in this work, with (blue) and without (green) a PEI. The bars indicate the value of the natural logarithm of the Bayesian evidence, normalized to Starobinsky inflation ($\mathrm{SI}$) without PEI, where the Jeffreys' scale is displayed with the vertical dotted lines for indication. The numerical values of $\ln(\mathcal{E}/\mathcal{E}_{\mathrm{SI}})$ are also explicitly written (in the case of $\mathrm{MSSMI}_\mathrm{o}$ and $\mathrm{RIPI}_\mathrm{o}$, they are smaller than the numerical accuracy of the \texttt{ASPIC} pipeline and only an upper bound is given). The best fit values are also shown with the black vertical ticks that are attached to left-pointing arrows, which stand for upper bounds on the Bayesian evidence for all possible priors. 
\label{fig:evidence}}
\end{center}
\end{figure}
Let us now include the models introduced in \Sec{sec:models} in the Bayesian pipeline of \texttt{ASPIC}. In \Fig{fig:evidence}, their Bayesian evidence is given, together with the maximal value of the likelihood over their prior space, $\umax_{\theta_{ij}}\mathcal{L} \left(\mathcal{D}\vert\theta_{ij},\mathcal{M}_i\right)$, \ie the ``best fit'' values. By definition, the best fit can only increase when a PEI is allowed since the parameter space extends. The Bayesian evidences have been normalized with respect to $\mathrm{SI}$ without PEI for reference but the normalization choice is irrelevant. What matters is the change in the relative Bayesian evidences when allowing the PEI, which shows how PEI mechanisms can substantially reorder the ranking of inflationary models. The most striking change concerns inflection point models ($\mathrm{MSSMI}_\mathrm{o}$ and $\mathrm{RIPI}_\mathrm{o}$) that are very strongly disfavored without a PEI but become almost weakly disfavored only when the PEI is allowed. We now analyze the different classes of models listed in \Sec{sec:models} in more details.
\subsection{Plateau models}
\label{sec:result:plateau}

As expected in \Sec{sec:model:plateau}, the Bayesian evidence of the Starobinsky model ($\mathrm{SI}$), representative of plateau potentials that already provide a good fit to the data without a premature end of inflation, decreases when the PEI is allowed, by a amount $\Delta \ln \mathcal{E}\simeq -2$. This is because the PEI explores regions of the potential that provide too large values for $\nS$. In the top left panel of \Fig{fig:MR:posterior}, the posterior distribution on the mass scale $M_\mathcal{R}$ associated with the curvature of the field space in the geometrical destabilization is displayed and translated into a posterior distribution on $\epsilon_\uc$. Since a PEI is disfavored, one obtains a lower bound on $M_\mathcal{R}$ that reads $\log_{10}(M_\mathcal{R}/\Mp) > -2.94$ at the two-sigma confidence level, which translates into $\epsilon_\uc>3.3\times 10^{-5}$.
\subsection{Large-field models}
\label{sec:result:largefield}
\begin{figure}[!ht]
\begin{center}
\includegraphics[width=0.328\textwidth]{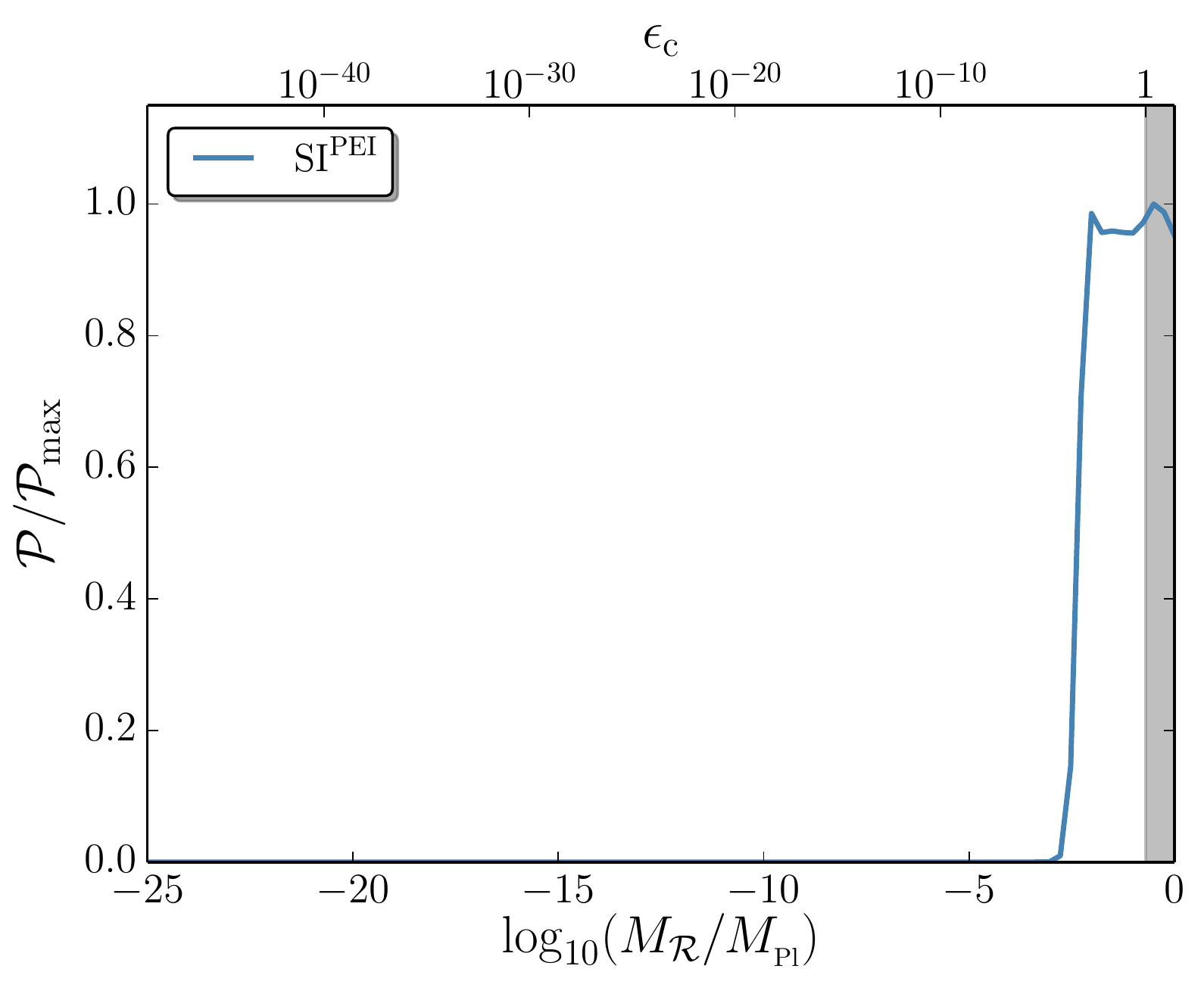}
\includegraphics[width=0.328\textwidth]{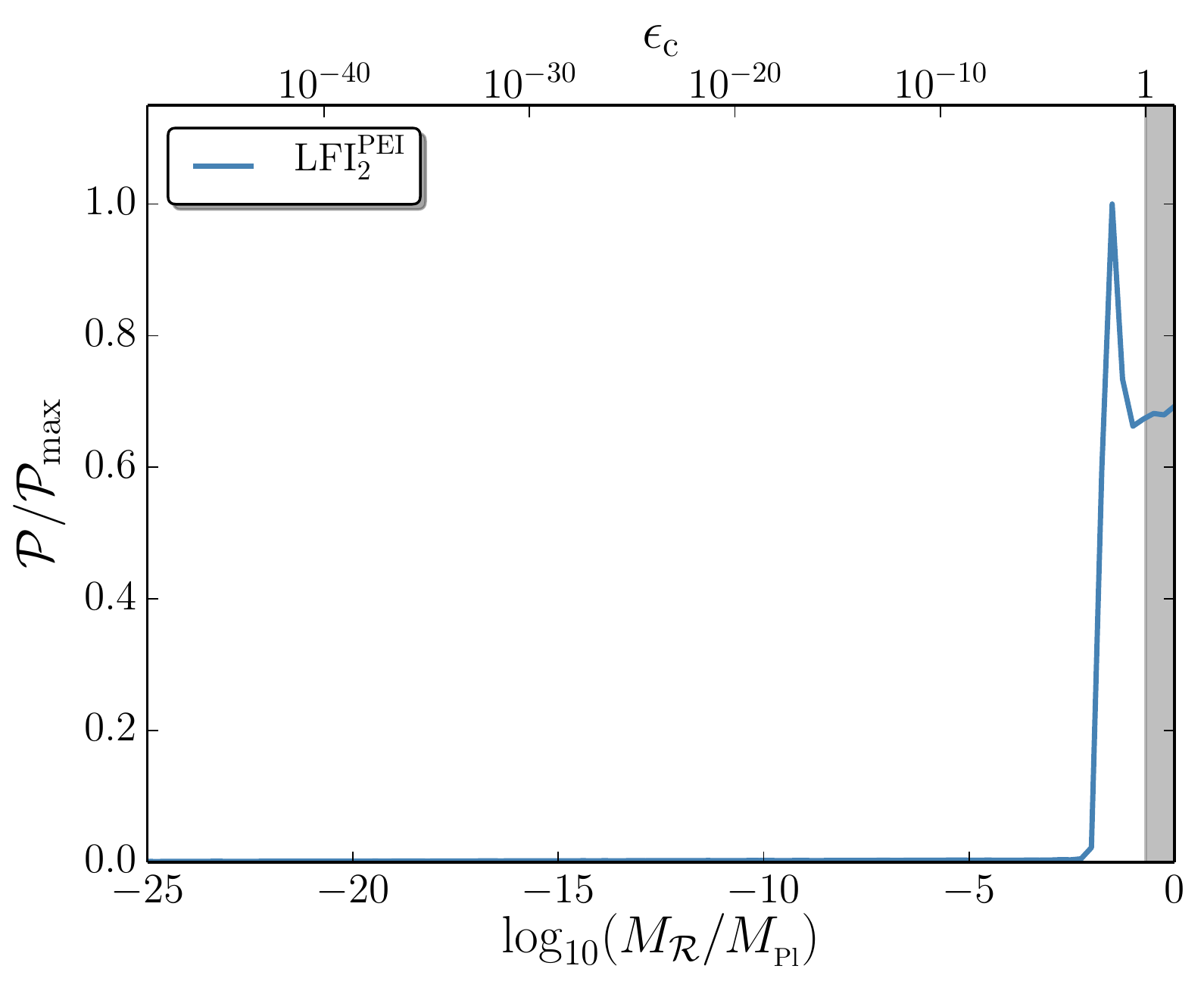}
\includegraphics[width=0.328\textwidth]{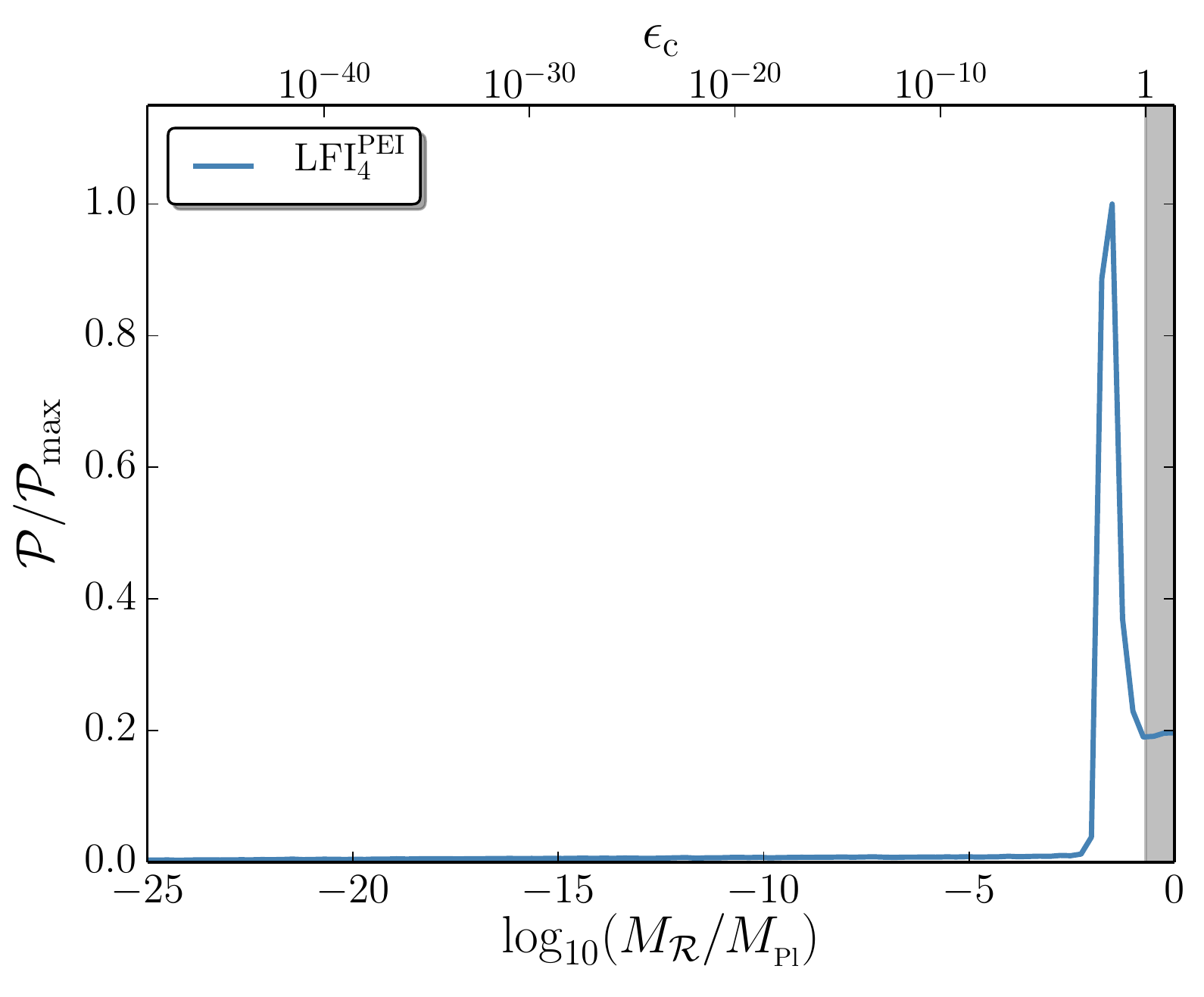}
\includegraphics[width=0.328\textwidth]{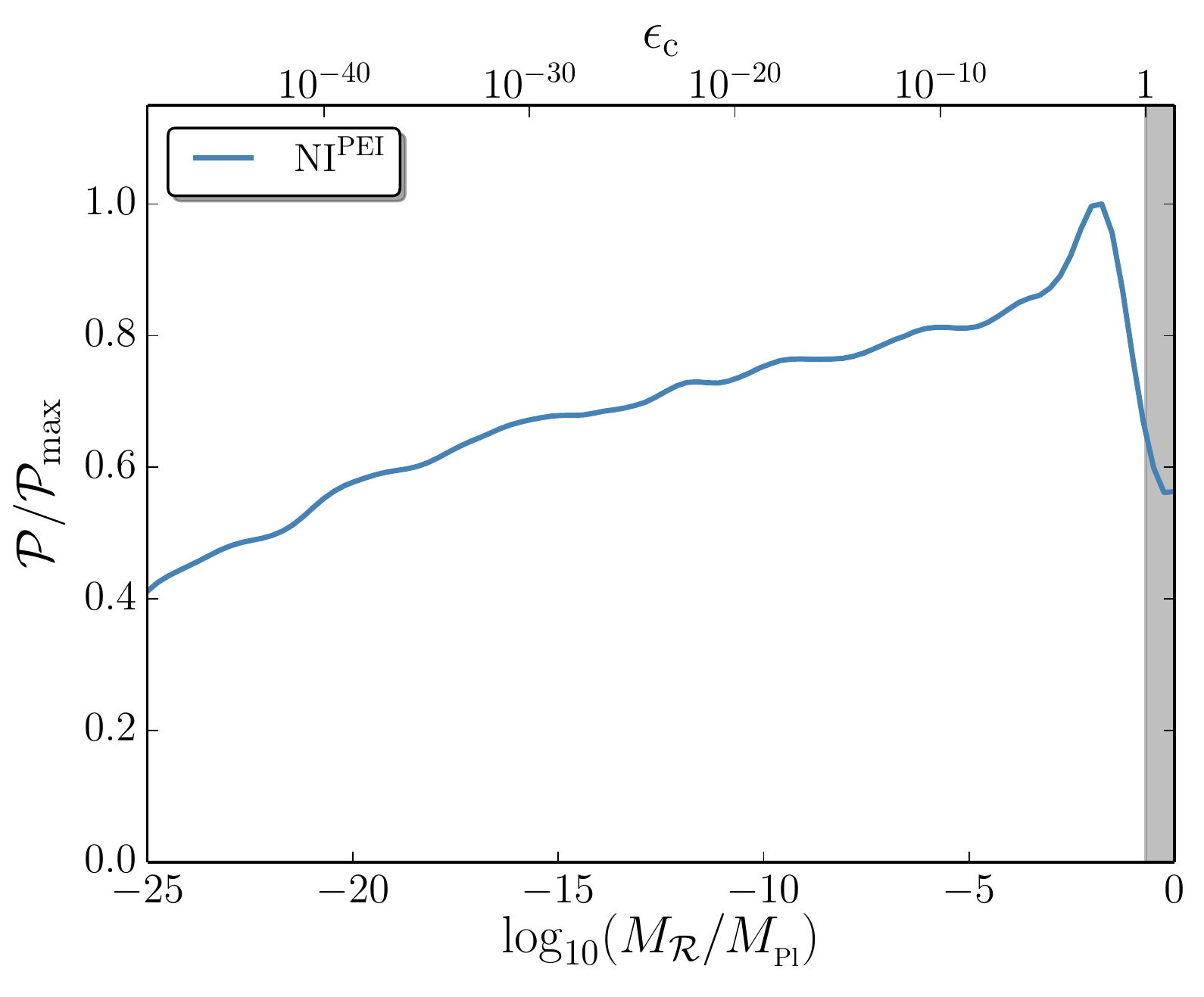}
\includegraphics[width=0.328\textwidth]{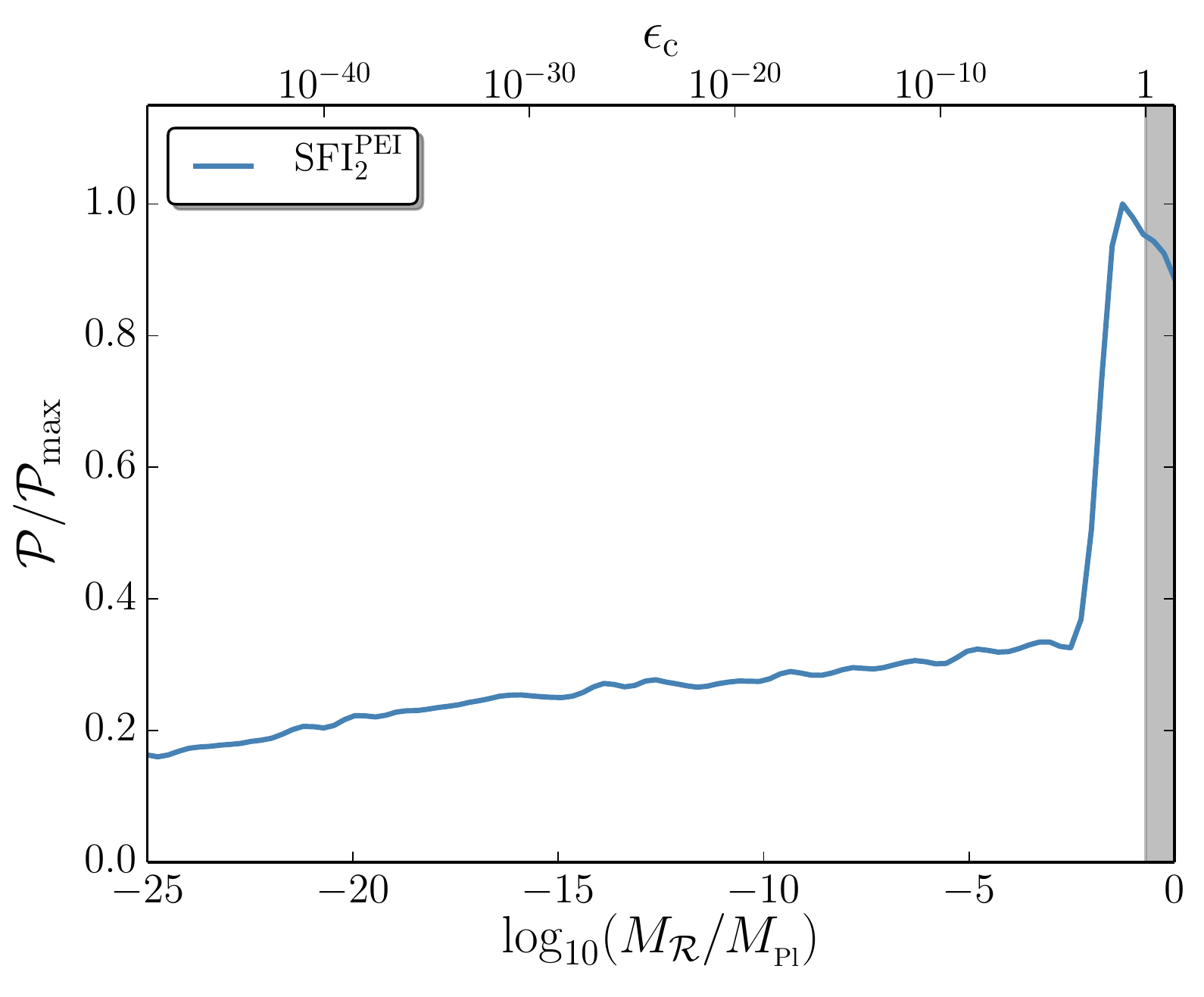}
\includegraphics[width=0.328\textwidth]{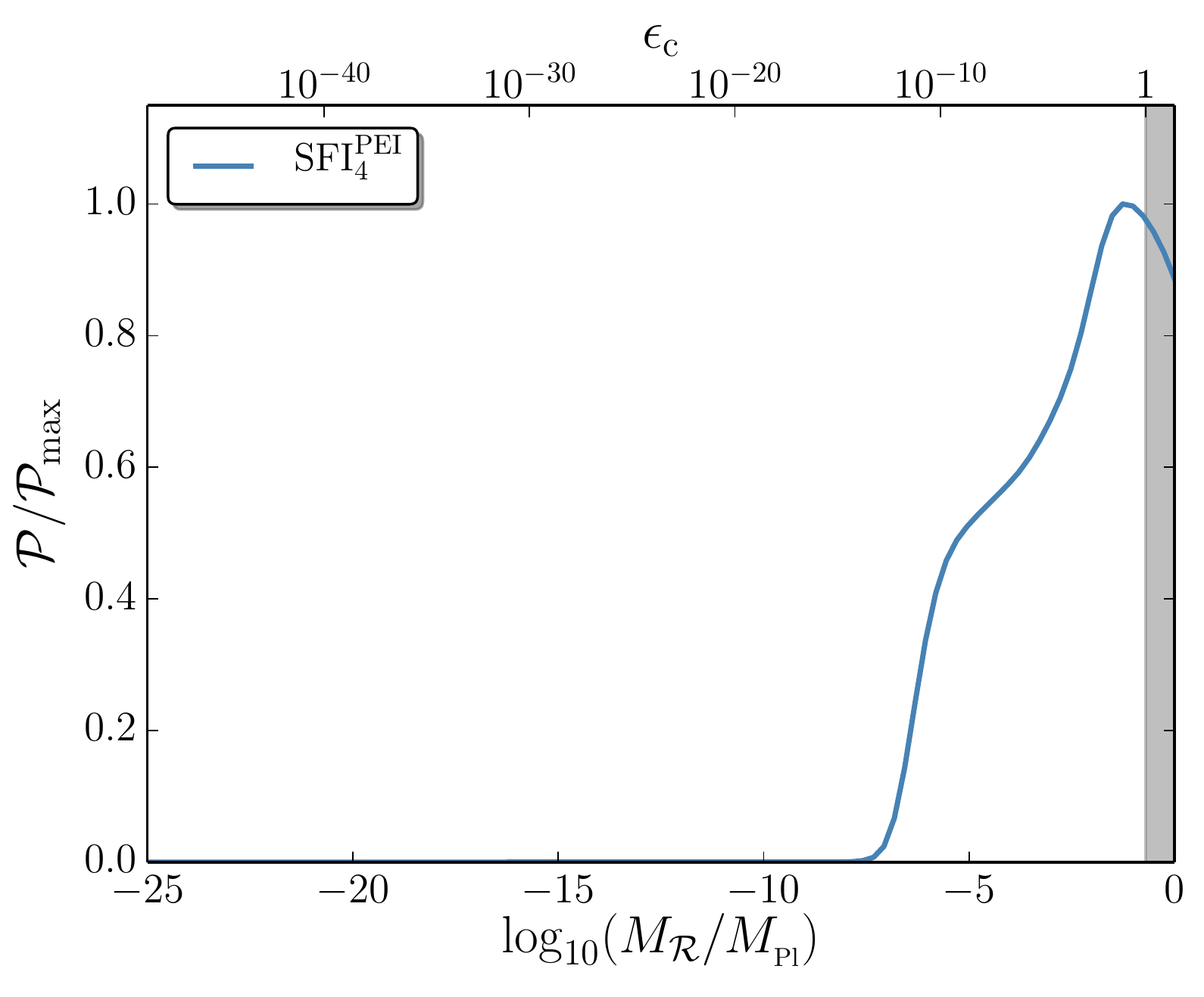}
\includegraphics[width=0.328\textwidth]{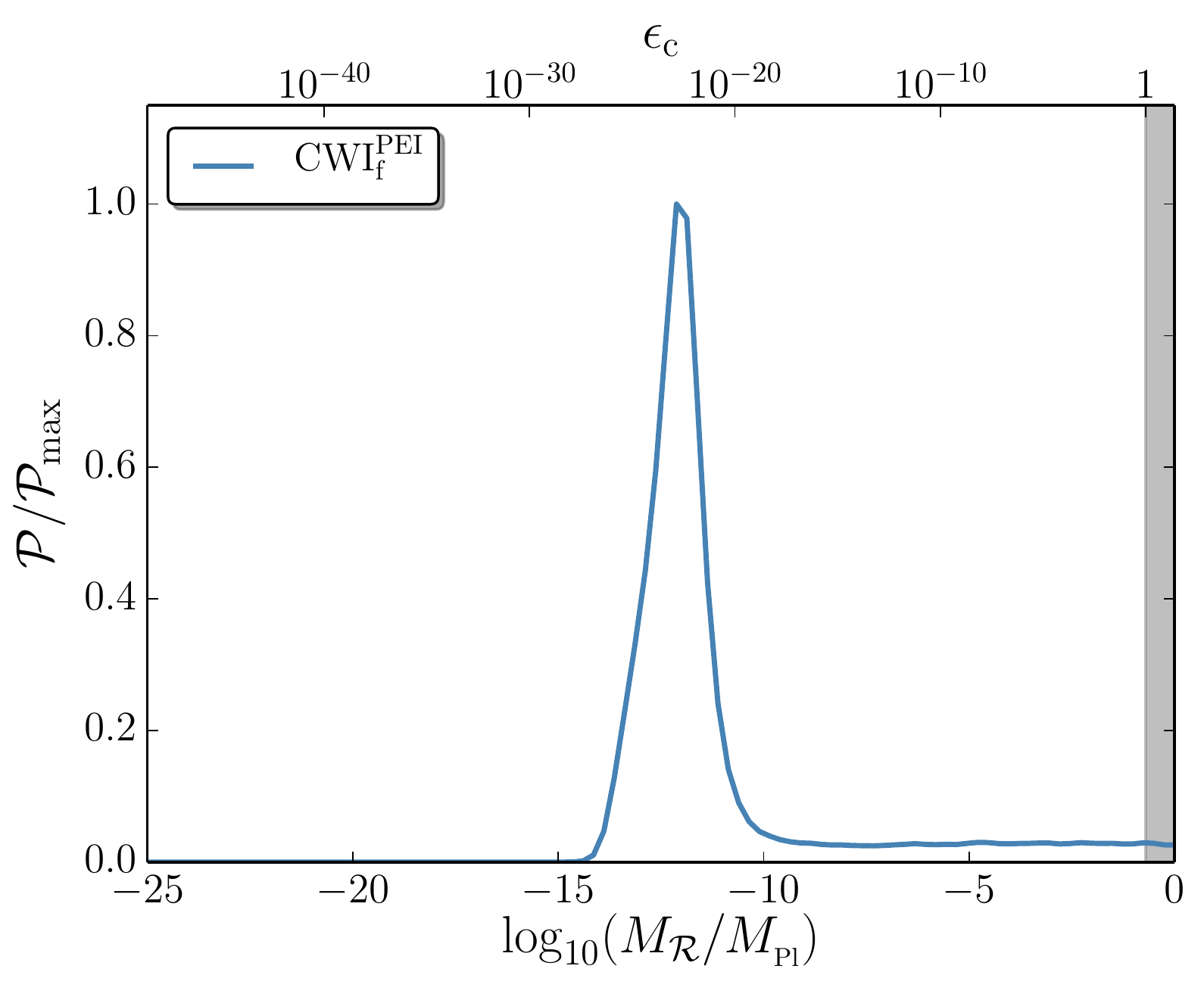}
\includegraphics[width=0.328\textwidth]{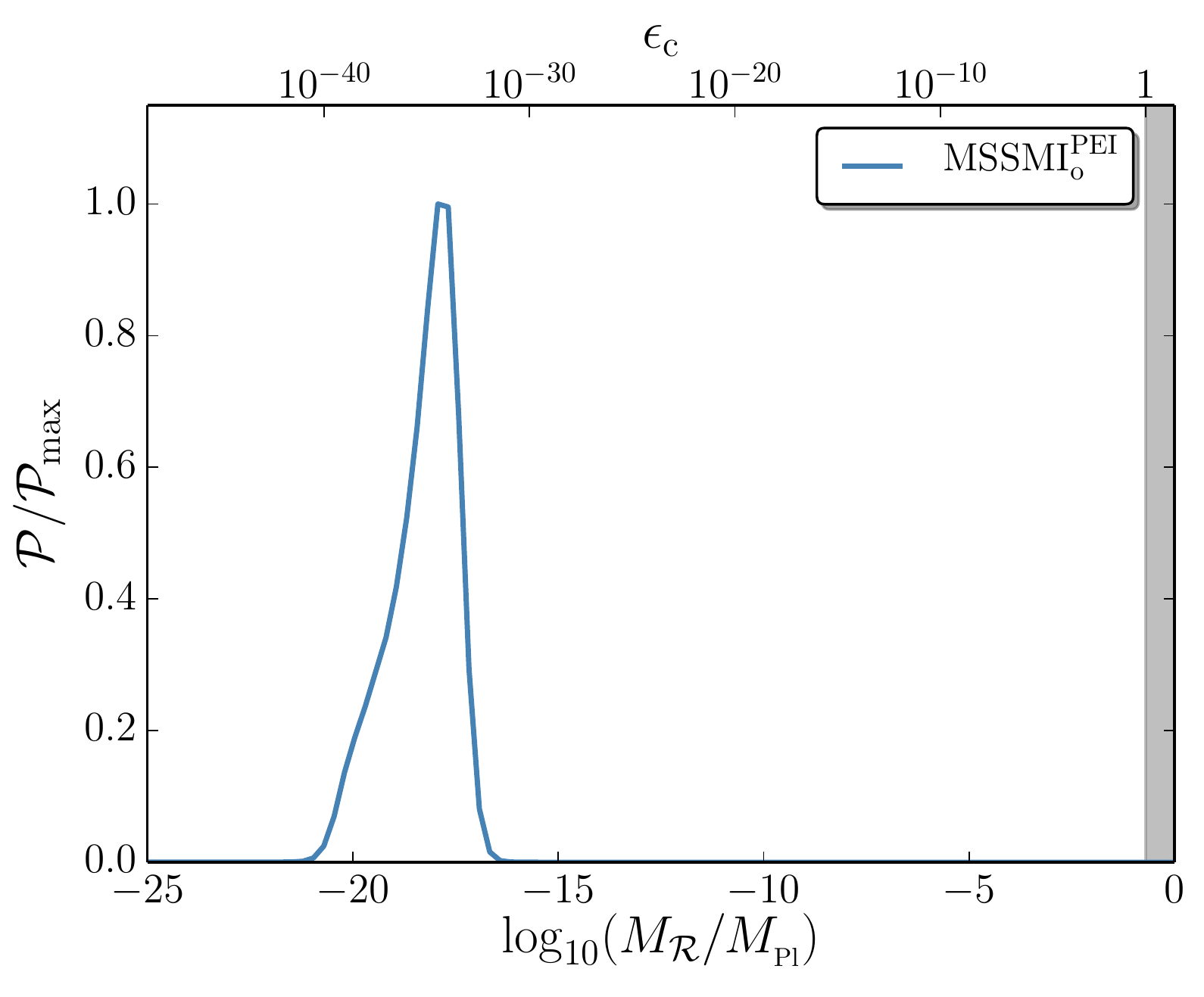}
\includegraphics[width=0.328\textwidth]{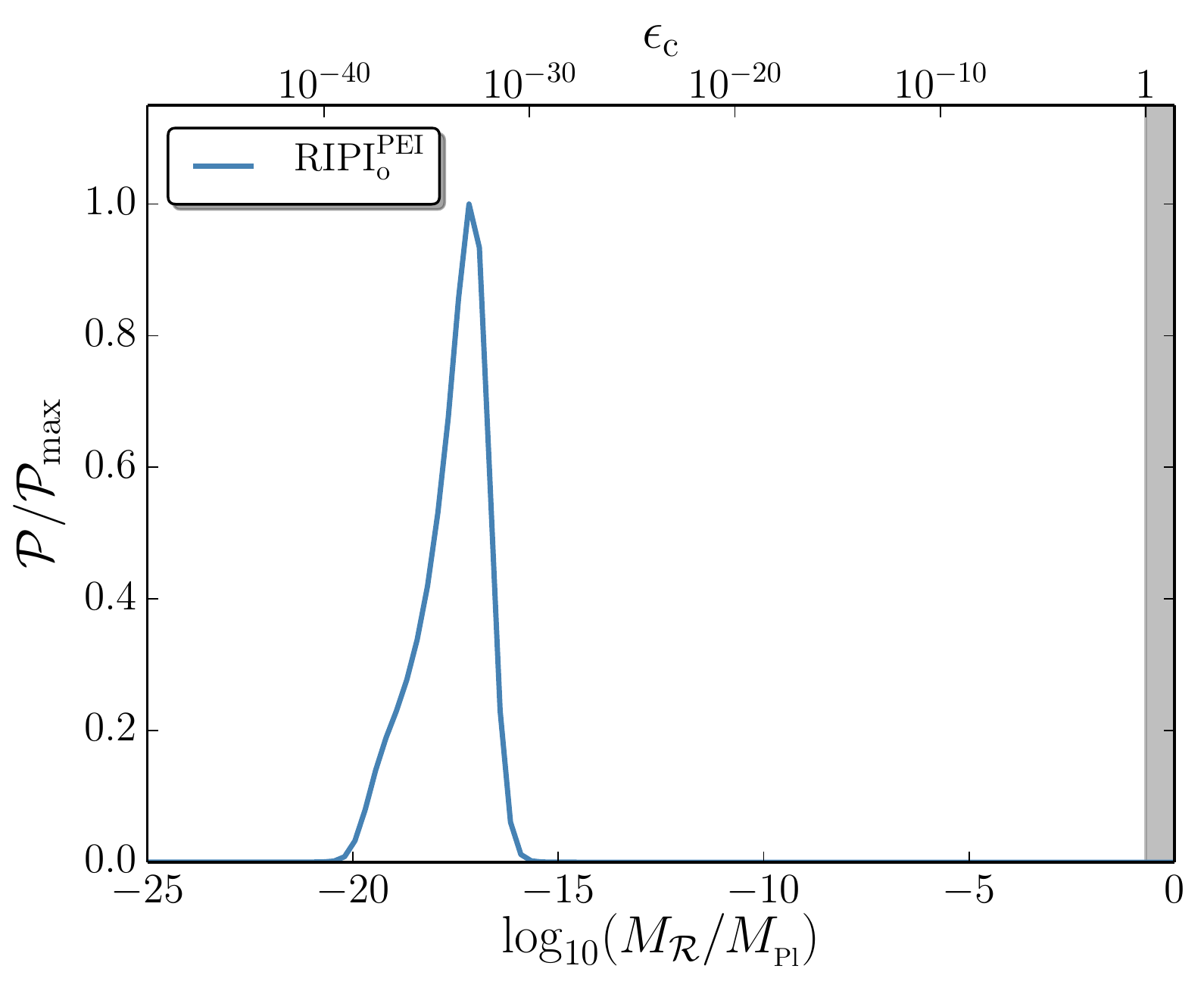}
\caption{Posterior distributions (normalized by their maximal values) on the mass scale $M_\mathcal{R}$ associated with the field space curvature in the geometrical destabilization and responsible for the premature end of inflation (PEI), for the models considered in this work. Making use of \Eq{eq:epsilonc}, since we have assumed $m_h/H_\uc=10$, each value of $M_\mathcal{R}$ translates into a value of $\epsilon_\uc$ that is labeled in the top axes. In this sense, these distributions can be seen as generic posteriors on $\epsilon_\uc$ regardless of the actual  mechanism ending inflation. The grey shaded areas stand for values of $\epsilon_\uc$ larger than one, where premature end of inflation does not occur and inflation ends by slow-roll violation as in the standard setup.
\label{fig:MR:posterior}}
\end{center}
\end{figure}

For large-field models that predict too large values of $r$ in the standard setup, here represented by $\mathrm{LFI}_2$ and $\mathrm{LFI}_4$, the Bayesian evidence decreases when a PEI is allowed, and both models become strongly disfavored. As explained in \Sec{sec:model:largefield}, this is because, even though the PEI allows smaller values of $r$ to be obtained, it is at the expense of larger values of $\nS$ that are even more disfavored by the data. 
However, as can be seen in the middle and top panels of \Fig{fig:nsr:prior}, the prior in the $(\nS,r)$ plane in these models comes closer to the observational contours with a PEI than without, with an improvement that is more pronounced for $\mathrm{LFI}_4$ than for $\mathrm{LFI}_2$, which is clearly visible in the best fits values of \Fig{fig:evidence}.
This explains why, in the middle and top panels of \Fig{fig:MR:posterior}, the posterior distribution on $M_\mathcal{R}$ peaks at intermediate values, namely $\log_{10}(M_\mathcal{R}/\Mp)=-1.52$ for $\mathrm{LFI}_2^\mathrm{PEI}$, corresponding to $\epsilon_\uc = 0.023$, and $\log_{10}(M_\mathcal{R}/\Mp)=-1.67 $ for $\mathrm{LFI}_4^\mathrm{PEI}$, corresponding to $\epsilon_\uc =0.012 $, where the peak is even more pronounced for $\mathrm{LFI}_4^\mathrm{PEI}$. However, because these peaks are very narrow, the values of $M_\mathcal{R}$ leading to an improvement of the fit are fine tuned and this explains why the Bayesian evidences decrease.

\subsection{Hilltop models with a non-vanishing mass at the top}
\label{sec:results:hilltopmass}
The Bayesian evidences of the hilltop models with a non-vanishing mass at the top of the hill is not strongly affected by the introduction of a PEI. $\mathrm{NI}$ slightly improves and $\mathrm{SFI}_2$ slightly worsens, but both models remain weakly disfavored with respect to $\mathrm{SI}$. As noted in \Sec{sec:models:hilltopmass}, this can be understood by examining the induced priors on $\nS$ and $r$ displayed in the left and middle panels of the second row of \Fig{fig:nsr:prior}, where one can see that for values of $r$ larger than $\sim 10^{-2}$, the two priors (with and without PEI) roughly coincide, at least at the one-sigma level, while for smaller values of $r$, they scan disjoint, but both disfavored, parameter space regions (namely values of $\nS$ that are too large with the PEI and too low without). This is why a premature end of inflation does not change much the Bayesian status of these models.

This is confirmed by studying the posterior distributions on $M_\mathcal{R}$ in the left and middle panels of the second row in \Fig{fig:MR:posterior}. In both cases, the distributions are rather flat, which is again consistent with the small impact PEI has on these models. For $\mathrm{NI}^\mathrm{PEI}$, the distribution slightly peaks at an intermediate value of $M_\mathcal{R}$, namely $\ln_{10}(M_\mathcal{R}/\Mp)= -2.79$, corresponding to $\epsilon_\uc = 6.7\times 10^{-5}$, and this peak is responsible for the slight increase in the Bayesian evidence. For $\mathrm{SFI}_2^\mathrm{PEI}$ however, the posterior distribution is maximal around the standard value $\epsilon_\mathrm{\uc}\simeq 1$, which explains why the Bayesian evidence decreases when $\epsilon_\uc$ is allowed to vary.

Another quantity of interest is the field value characterizing the width of the hill, namely $f$ for $\mathrm{NI}$ and $\mu$ for $\mathrm{SFI}_2$. The posterior distributions on these parameters is displayed in \Fig{fig:Vparam:posterior}. In the absence of a PEI, one can check that only super-Planckian values of these parameters are allowed by the data, since values of the order or smaller than the Planck mass lead to values of $\nS$ that are too small. Generating a super-Planckian hill width in a consistent complete UV theory is not an easy task in these models and has been the subject of an abundant literature. One can see that allowing a PEI does not alleviate this problem since Planckian or sub-Planckian values of $f$ and $\mu$ are still strongly disfavored. However, it removes the super-Planckian tail of the distributions and leads to a clear measurement of these parameters, namely $\log_{10}(f/\Mp)=0.77\pm0.24$ for $\mathrm{NI}^\mathrm{PEI}$ and $\log_{10}(\mu/\Mp)=1.08\pm 0.19$ for $\mathrm{SFI}_2^\mathrm{PEI}$ at the one-sigma level.\footnote{The difference between the two mean values is of order $\log_{10}(2)$, which corresponds to the relationship between $f$ and $\mu$ obtained by Taylor expanding the two potentials and identifying them as explained below \Eq{eq:pot:sfi2}. This means that the region of interest in the parameter space is such that $\phi_*\ll \mu$ or $\phi_*\ll f$, where the two potentials are indeed approximately the same. This is consistent with the fact that they have very comparable Bayesian evidence when the PEI is allowed.}
\subsection{Hilltop models with a vanishing mass at the top}
Hilltop models with a vanishing mass at the top of the hill are substantially more affected by PEI mechanisms, although the effect depends on the details of the potential one considers.

\begin{figure}[!ht]
\begin{center}
\includegraphics[width=0.328\textwidth]{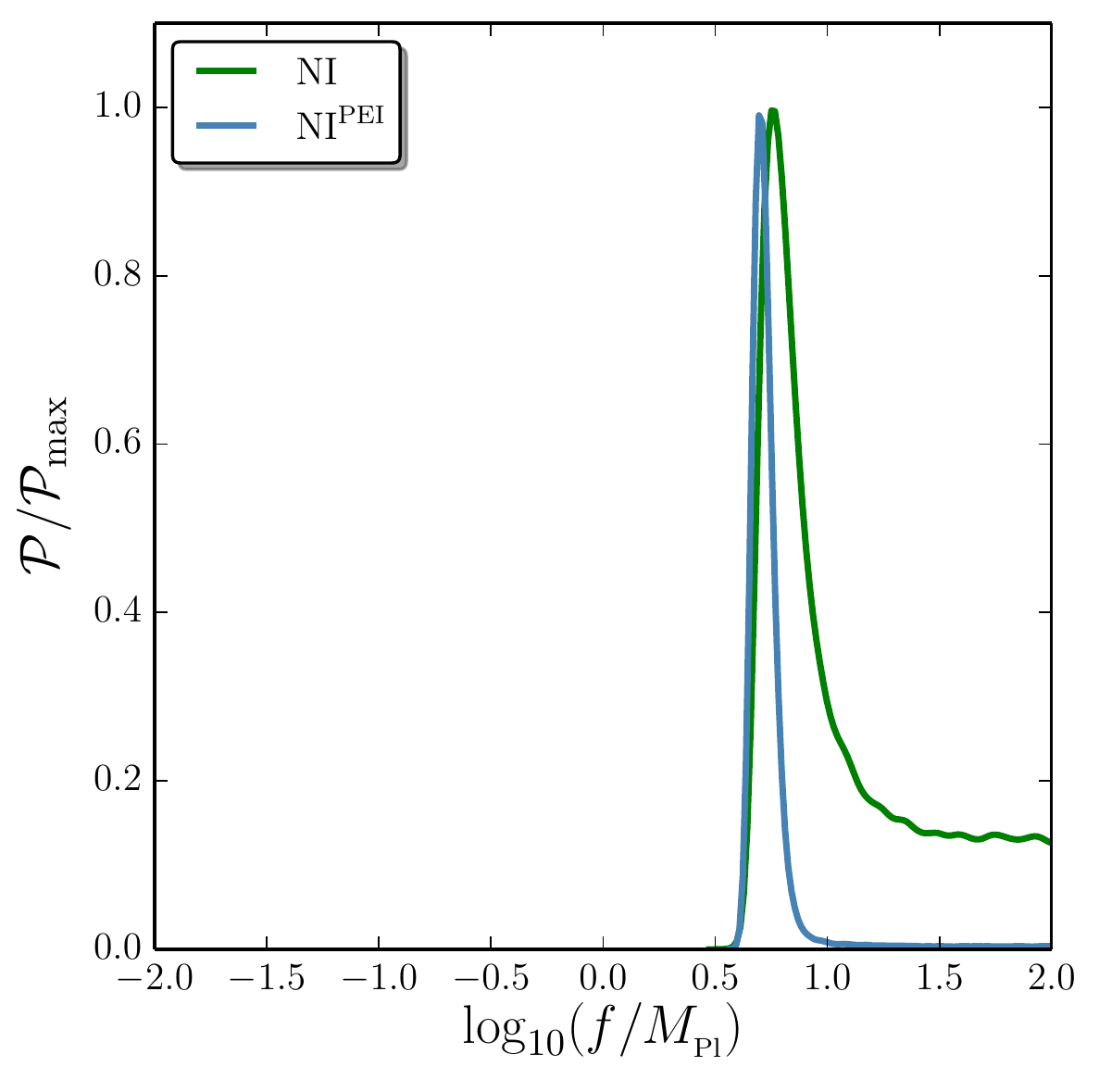}
\includegraphics[width=0.328\textwidth]{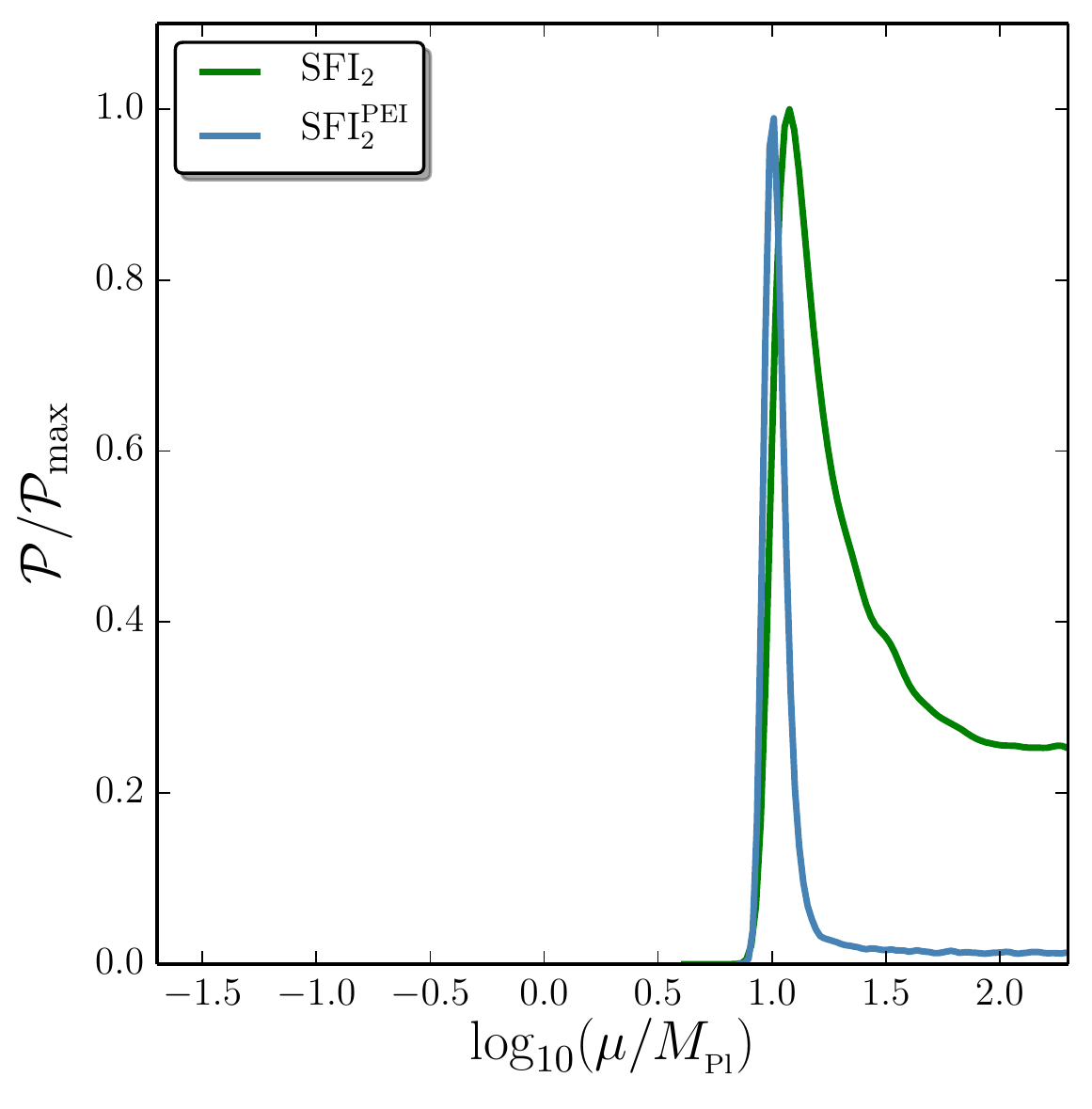}
\includegraphics[width=0.328\textwidth]{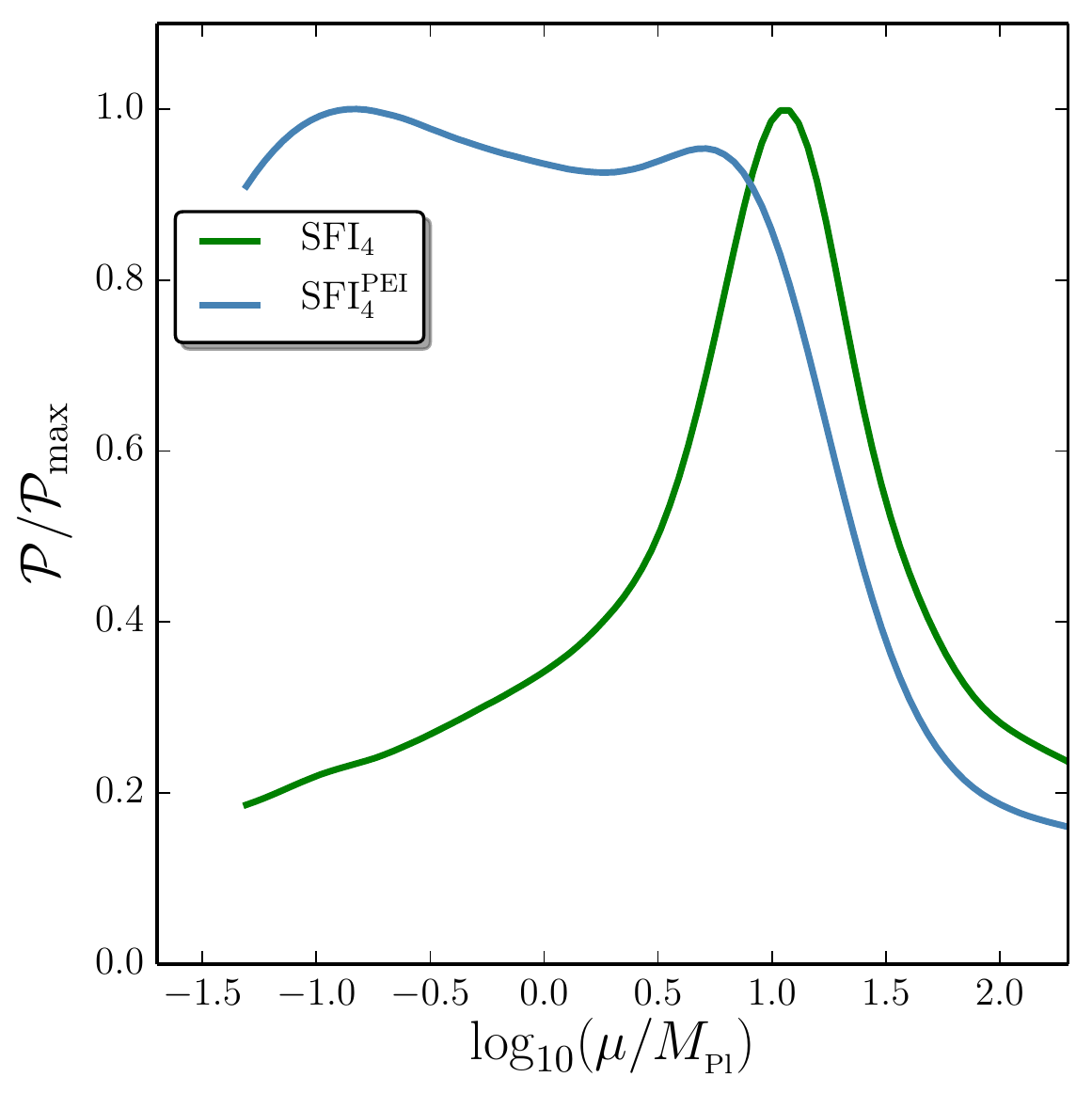}
\includegraphics[width=0.328\textwidth]{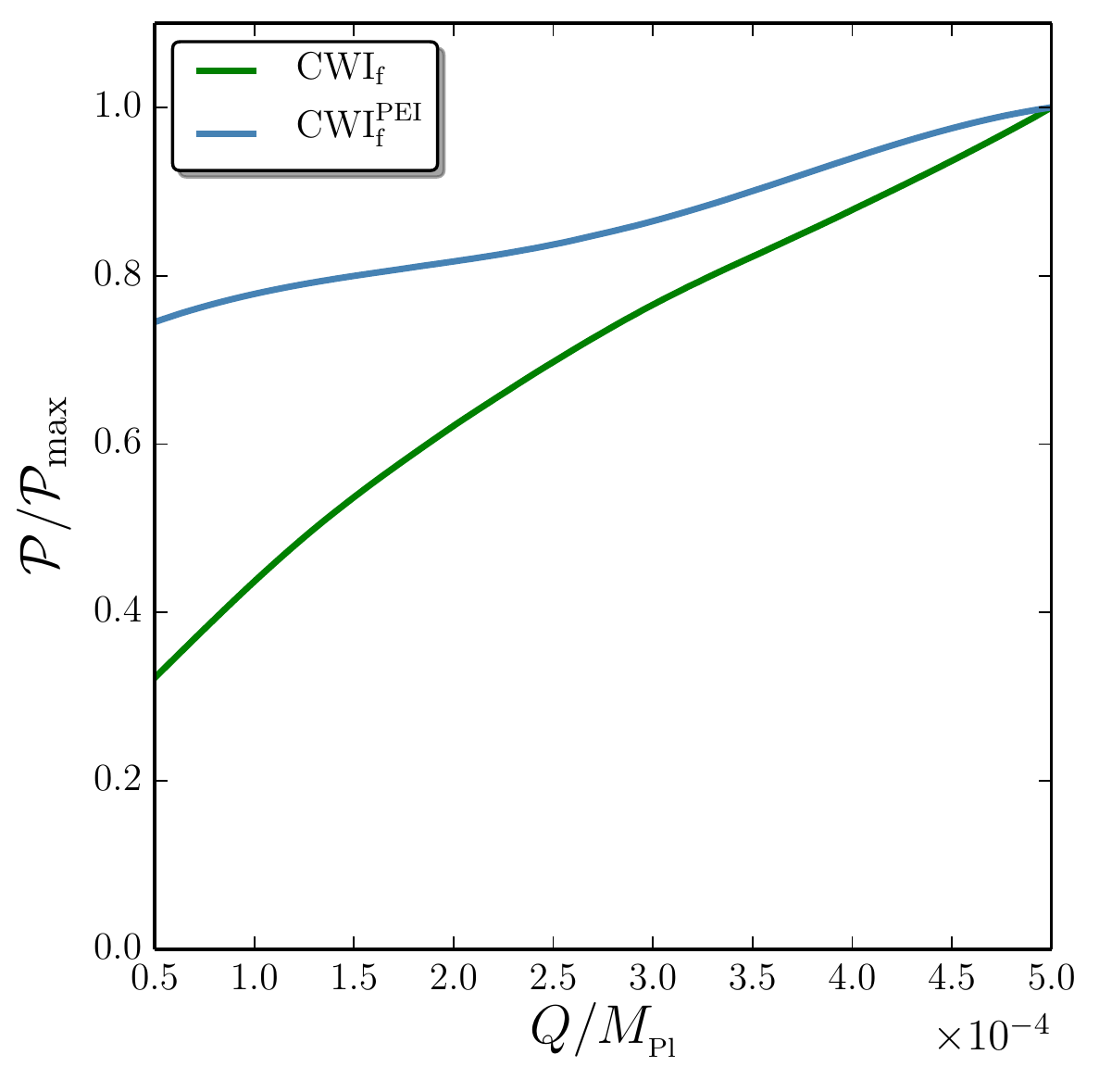}
\includegraphics[width=0.328\textwidth]{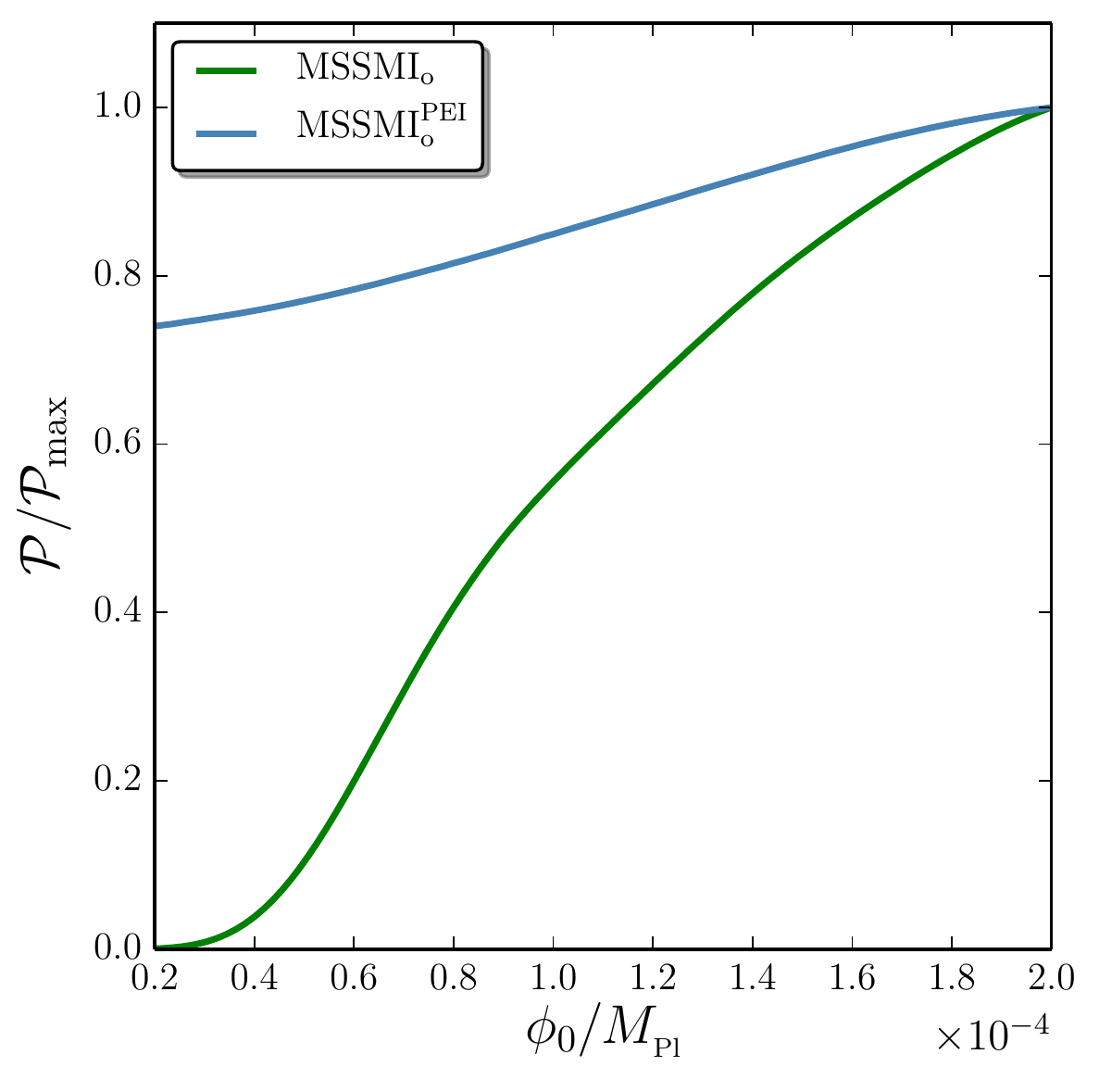}
\includegraphics[width=0.328\textwidth]{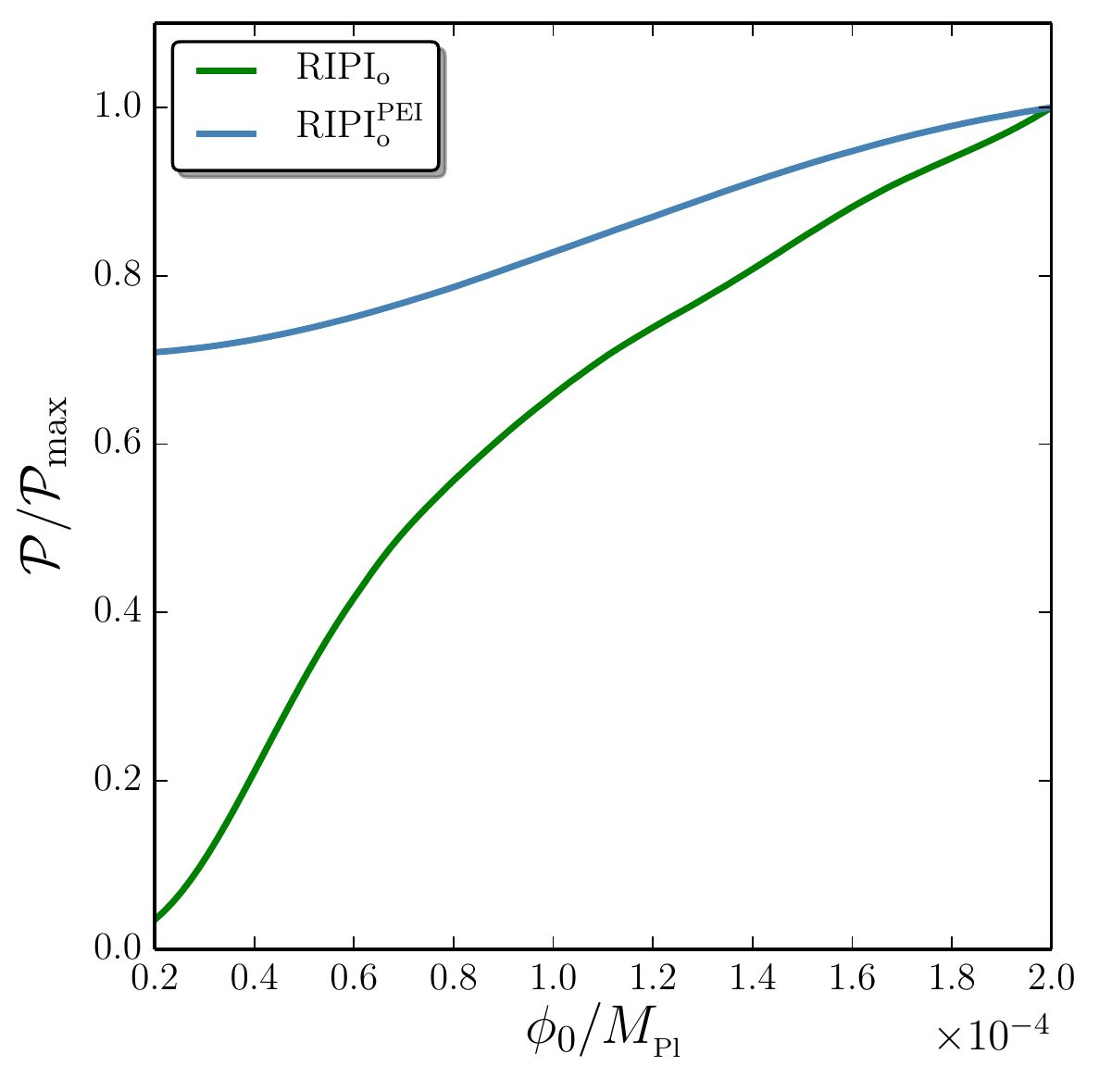}
\caption{Posterior distributions on the parameters of the potentials studied in this work, with (blue) and without (green) premature end of inflation (PEI).
\label{fig:Vparam:posterior}}
\end{center}
\end{figure}

For $\mathrm{SFI}_4$, the PEI decreases the Bayesian evidence of the model, which becomes weakly disfavored. This is because the spectral index $\nS$, which is slightly too small (but still compatible with the data) in the standard setup, is mostly predicted to be close to scale invariance when the PEI is allowed, as can be seen in the one-sigma contours of the right panel of the third row of \Fig{fig:nsr:prior}, which is observationally excluded. This is why, contrary to $\mathrm{SFI}_2$ discussed in \Sec{sec:results:hilltopmass}, small values of $M_\mathcal{R}$ are excluded as can be seen in the posterior distribution of $M_\mathcal{R}$ displayed in the middle right panel of \Fig{fig:MR:posterior}. More precisely, one gets the two-sigma constraint $\log_{10}(M_\mathcal{R}/\Mp)>-7.8$, corresponding to $\epsilon_\mathrm{c}>4.5\times 10^{-6}$. As for $\mathrm{SFI}_2$, it is also worth discussing the posterior distribution on $\mu$, displayed in the top right panel of \Fig{fig:Vparam:posterior}. In the absence of a PEI, and contrary to $\mathrm{SFI}_2$, one can see that sub-Planckian values of $\mu$ are marginally allowed since they give rise to values of $\nS$ that are not excluded by the data. However, the model still features some preference for Planckian or slightly super-Planckian values of $\mu$ or order $\mu\sim 10\Mp$. When the PEI is allowed however, the sub-Planckian tail of the distribution is lifted up to a plateau which overall shows preference for sub-Planckian values of $\mu$. This is in sharp contrast with $\mathrm{SFI}_2$ and sheds new light on the problem of getting super-Planckian hill widths since in $\mathrm{SFI}_4^\mathrm{PEI}$, this is not a requirement anymore. Let us however note that this is at the expense of making the model weakly disfavored overall.

For $\mathrm{CWI}_{\mathrm{f}}$, moderately disfavored in the standard case since it predicts values of $\nS$ that are too low, the model improves when a PEI is allowed and becomes weakly disfavored only. From the induced prior on $\nS$ and $r$ displayed in the bottom left panel of \Fig{fig:nsr:prior}, we already noted in \Sec{sec:models:hilltop:massless} that with a PEI, either $\nS$ is predicted to be at the level obtained in the standard setup, which is too low, or it is predicted to be close to scale invariance, which is too large. This is why a PEI does not fully succeed in making the model favored. In between these two peaks of the prior distribution, the predictions sweep the data's sweet spot and this is why intermediate values of $M_\mathcal{R}$ are strongly preferred in the posterior distribution displayed in the bottom left panel of \Fig{fig:MR:posterior}. It is interesting to notice that $M_\mathcal{R}$ is accurately measured in this model, and one obtains $\log_{10}(M_\mathcal{R}/\Mp)= -11.1  \pm 2.9 $ at the one-sigma confidence level, corresponding to $\log_{10}(\epsilon_\uc) = -20.7 \pm 7.2 $. The posterior distribution on $Q$, the parameter appearing in the potential~(\ref{eq:pot:cwi}), is displayed in the bottom left panel of \Fig{fig:Vparam:posterior} but is weakly constrained with or without PEI.

\afterpage{\FloatBarrier}
\subsection{Inflection point models}
\label{sec:results:inflection}
The situation of inflection point models is very similar to the one of $\mathrm{CWI}_\mathrm{f}$ but even more pronounced, since the values predicted for $\nS$ are even smaller without PEI than the ones for $\mathrm{CWI}_\mathrm{f}$. This explains why the shift in the Bayesian evidence of $\mathrm{MSSMI}_\mathrm{o}$ and $\mathrm{RIPI}_\mathrm{o}$ is even larger than the one observed for $\mathrm{CWI}_\mathrm{f}$, even though the Bayesian evidences of these three models with PEI is comparable. As for $\mathrm{CWI}_\mathrm{f}$, the posterior distribution of $M_\mathcal{R}$ for $\mathrm{MSSMI}_\mathrm{o}$ and $\mathrm{RIPI}_\mathrm{o}$, displayed in the bottom middle and right panels of \Fig{fig:MR:posterior}, has a sharp peak, which leads to a measurement of $M_\mathcal{R}$. One finds $\log_{10}(M_\mathcal{R}/\Mp) = -18.3\pm 0.8$ for $\mathrm{MSSMI}_\mathrm{o}$, corresponding to $\log_{10}(\epsilon_\uc) = -35.2\pm 3.1$, and $\log_{10}(M_\mathcal{R}/\Mp) =  -17.6 \pm 0.8$ for $\mathrm{RIPI}_\mathrm{o}$, corresponding to $\log_{10}(\epsilon_\uc) = - 33.8\pm 3.1$. The posterior distributions on $\phi_0$, the parameter appearing in the potentials~(\ref{eq:pot:mssmi}) and~(\ref{eq:pot:ripi}), are displayed in the bottom right panels of \Fig{fig:Vparam:posterior} but are  weakly constrained with or without GD. Eventually, it is interesting to translate the constraints on $M_\mathcal{R}$ on the derived parameter\footnote{Notice that $\log_{10}(M_{\mathcal{R}}/H_*)$ is not a parameter that is directly sampled in the present analysis, where one assumes a flat prior on $\log_{10}(M_{\mathcal{R}}/\Mp)$, $H_*$ is computed from the power spectrum normalisation, and $\log_{10}(M_{\mathcal{R}}/H_*)$ is obtained for each point in the chains, from which its posterior distribution is derived. Because the induced prior on this derived parameter is not flat, the one-sigma constraint quoted in the main text contains information not only from the data but also from the prior. However, the constraint is so sharp that the induced prior can be approximated as being almost constant on the one-sigma range, which is therefore essentially driven by the data and mildly depends on the prior. Notice that except from $\log_{10}(M_{\mathcal{R}}/H_*)$, all constraints quoted in this article are on quantities on which a flat prior is assumed.} $M_\mathcal{R}/H_*$, obtaining for both models $\log_{10}(M_\mathcal{R}/H_*) = 3.0\pm 0.2$.

A few comments are now in order. We studied two inflection point models only, but they are typical for models
of inflation arising in string theory \cite{Baumann:2007np,Krause:2007jk,Baumann:2007ah,Agarwal:2011wm,McAllister:2012am}. With $H_*\sim\mathcal{O}(\mathrm{MeV})$, these are models of low-scale inflation. 
Values of $M_\mathcal{R}\sim\mathcal{O}(\mathrm{GeV})$ may be considered
extremely low for a cutoff scale in the effective dimension-6 operator $(\partial\phi)^2\chi^2/M_\mathcal{R}^2$ from a particle physics point of view, but this is just another incarnation of extreme fine-tuning present in these models \cite{Lalak:2007rsa}. In fact, this scale of high-energy effects lies three orders of magnitude above the scale of inflation $H_*$ and it is quite remarkable that it can be constrained observationally without resorting to primordial non-Gaussianities.

\section{Discussion and conclusion}
\label{sec:concl}

Motivated by the mechanism of the geometrical destabilization of inflation, we have investigated in this work how the Bayesian ranking of single-field slow-roll models of inflation is affected when allowing a mechanism of premature ending of inflation. We have found that plateau potentials that already provide a good fit to the data can only be made worse in the presence of a premature termination of inflation, and that large-field models that lead to values of the tensor-to-scalar ratio $r$ that are too large in the standard setup are still disfavored when a PEI is allowed since the reduction of $r$ that it yields is to the detriment of a too large increase in the value of the scalar spectral index $\nS$. Quadratic hilltop models, that predict values of $\nS$ that are too low when the hill has a sub-Planckian width, are not largely affected by the PEI. This is because, even though the PEI increases the values of $\nS$, it allows the models to match observational constraints only in a fine-tuned range of the parameter space, and otherwise yields values of $\nS$ that are too large. Quartic hilltop models on the other hand can be more substantially affected by a PEI, in a way that however depends on the details of the potential. In the case of $\mathrm{SFI}_4$ where $V\propto 1-(\phi/\mu)^4$, contrary to the standard case, a PEI favors sub-Planckian values of $\mu$, that are more natural in these models. Finally, inflection point models which predict values of $\nS$ that are too low in the standard case and are therefore strongly disfavored, are only weakly disfavored when a PEI is allowed. In this case, and when interpreted in the framework of the geometrical destabilization, sharp measurements of the field-space curvature mass scale $M_\mathcal{R}$, at the level of the $\mathrm{GeV}$ scale, were derived. These results demonstrate how the interpretation of cosmological data in terms of fundamental physics and model building can be drastically modified in the presence of a premature end of inflation, as motivated by the mechanism of the geometrical destabilization.

By discussing a few classes of models, more involved behaviors can also be addressed by viewing our prototypical examples as building blocks for more complicated phenomenologies. For example, $\alpha$-attractor models~\cite{Kallosh:2013hoa, Kallosh:2013yoa}, which interpolate between the Starobinsky and large-field potentials, can be discussed by combining the results of \Secs{sec:result:plateau} and~\ref{sec:result:largefield}. Since both the Starobinsky model and the large-field potentials become worse in the presence of a PEI, $\alpha$-attractors are most certainly also worsened by allowing a PEI. Let us also note that the situation of models predicting a value for $\nS$ that is too large in the standard case has not been explicitly discussed so far. In fact, two cases can be distinguished, depending on whether the value predicted for $\nS$ is too large but still red, $\nS<1$, or blue, $\nS>1$. Not many models fall in the first category, the typical example being power-law inflation for which $V(\phi)\propto \exp(-\alpha\phi/\Mp)$. This potential is however conformally invariant so that changing the end of inflation location has exactly no impact on the predictions of the model. In the second case, $\nS>1$, $\epsilon$ decreases as inflation proceeds (this is because in single-field slow-roll inflation, $\nS\simeq 1- 2 \epsilon - \dd\ln\epsilon/\dd N$ and $\epsilon$ is always positive), which means that $m_{\mathrm{eff}}^2$ in \Eq{eq:meff} increases and the GD cannot take place. So a PEI has to be realized through another mechanism. In that case, a potential of the form $V\propto 1+(\phi/\phi_0)^p$, which predicts $\nS>1$ at $\phi\ll \phi_0$ and $\nS<1$ at  $\phi\gg \phi_0$, could be turned from blue to red if inflation ends prematurely. However, in the  $\phi\gg \phi_0$ regime, the model asymptotes large-field inflation, which has been shown to be disfavored with or without PEI in \Sec{sec:result:largefield}. Since the model interpolates between two disfavored limits, it is likely disfavored as well. 

\begin{figure}[!t]
\begin{center}
\includegraphics[width=0.328\textwidth]{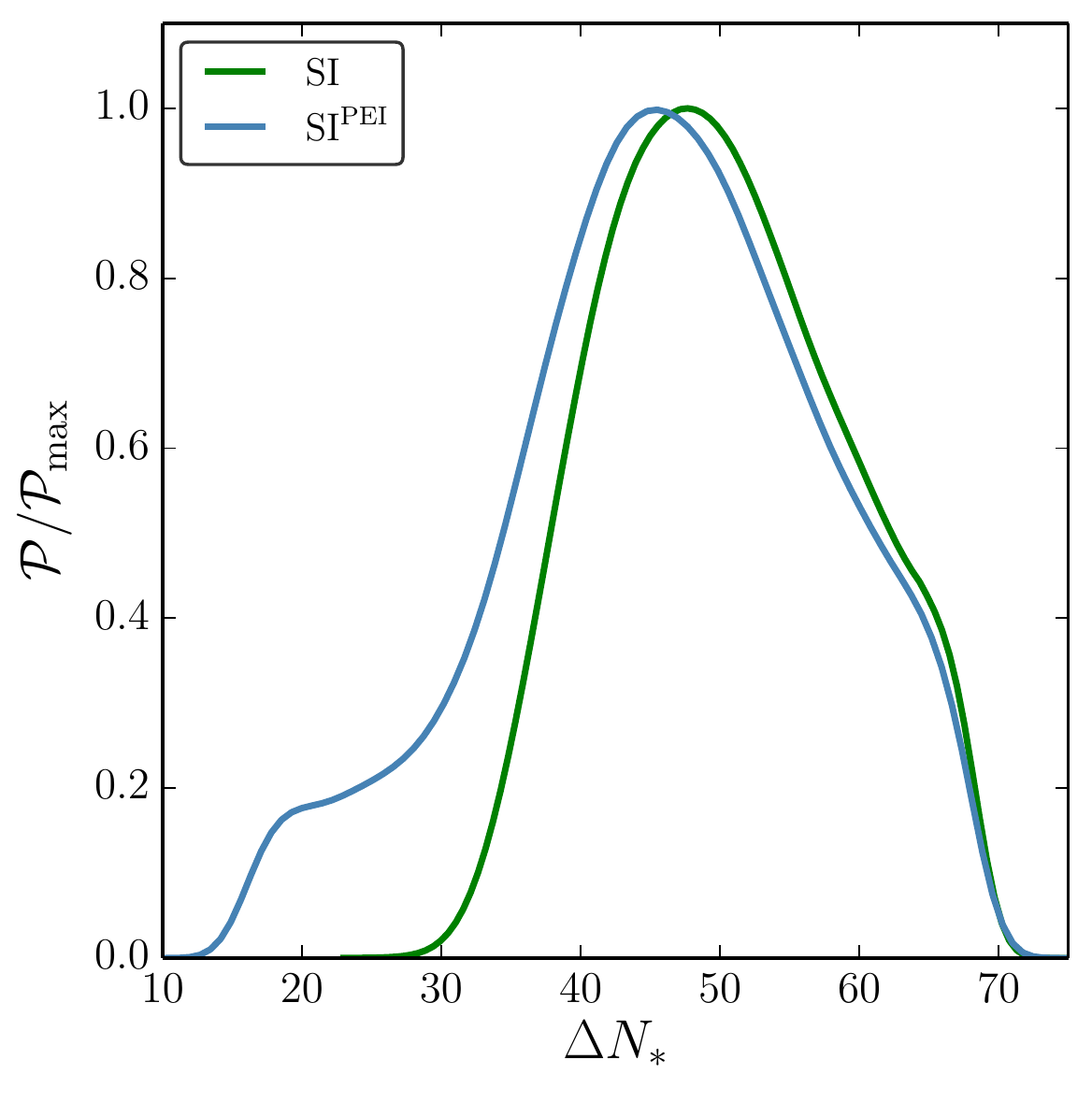}
\includegraphics[width=0.328\textwidth]{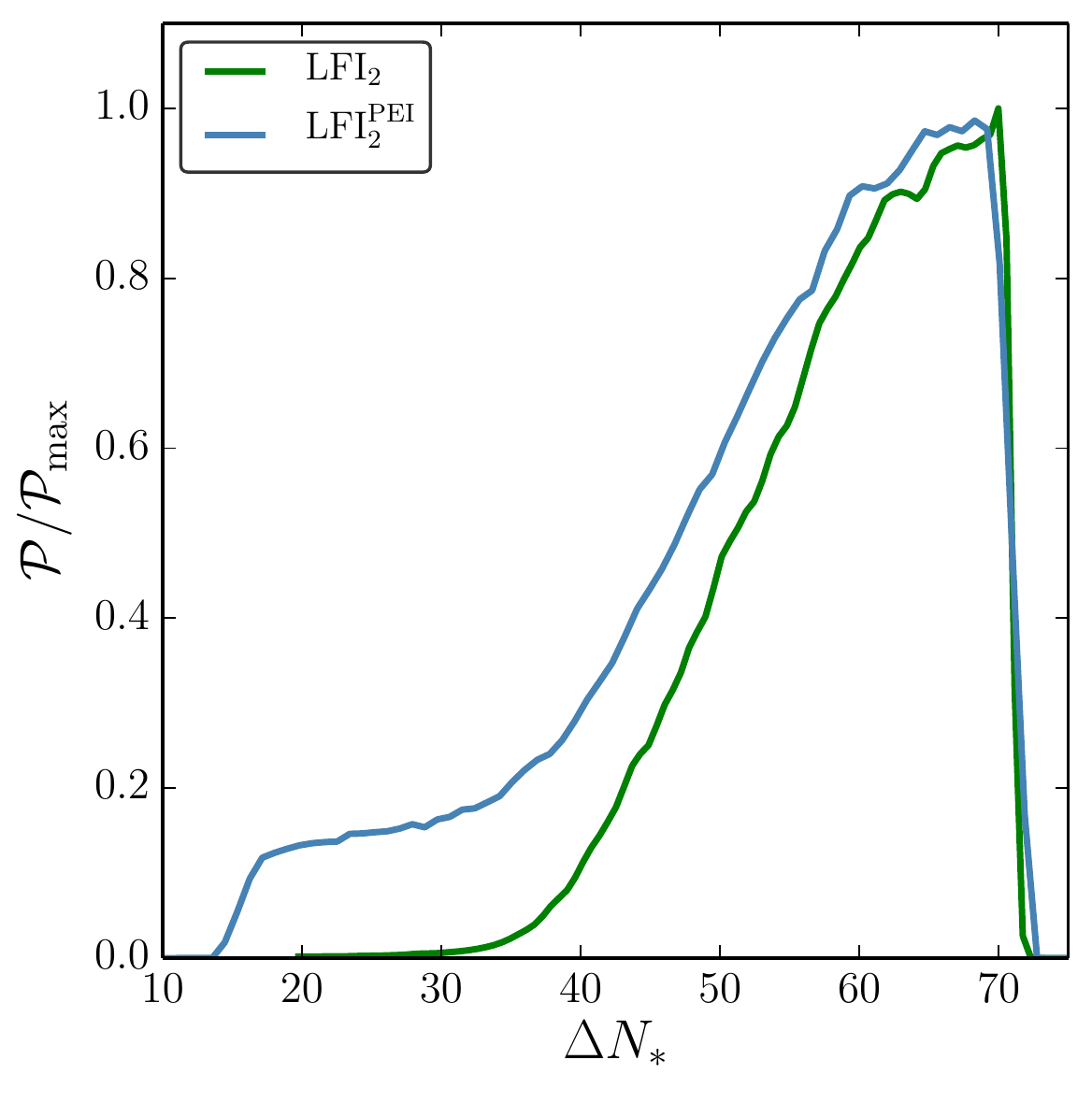}
\includegraphics[width=0.328\textwidth]{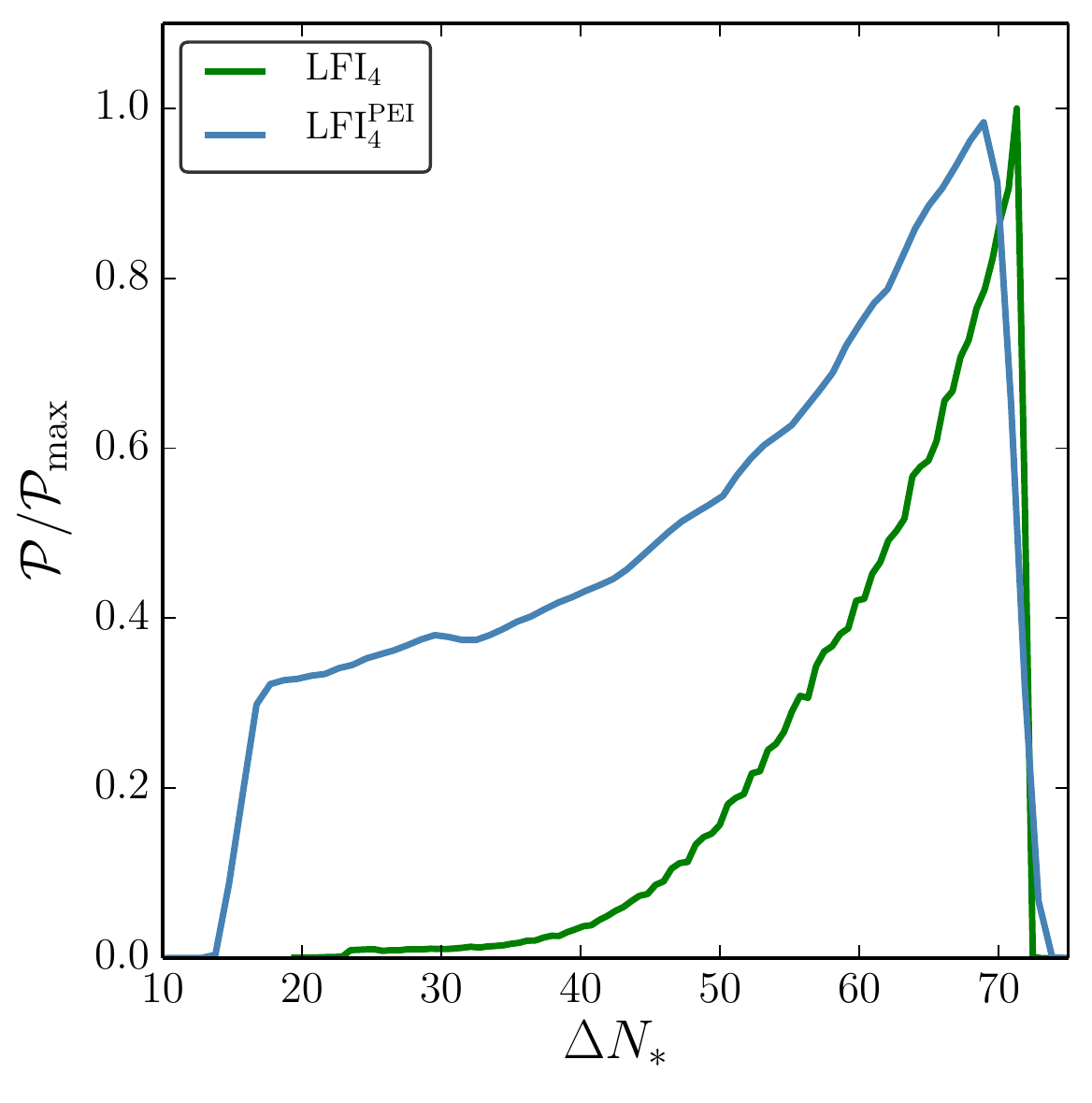}
\includegraphics[width=0.328\textwidth]{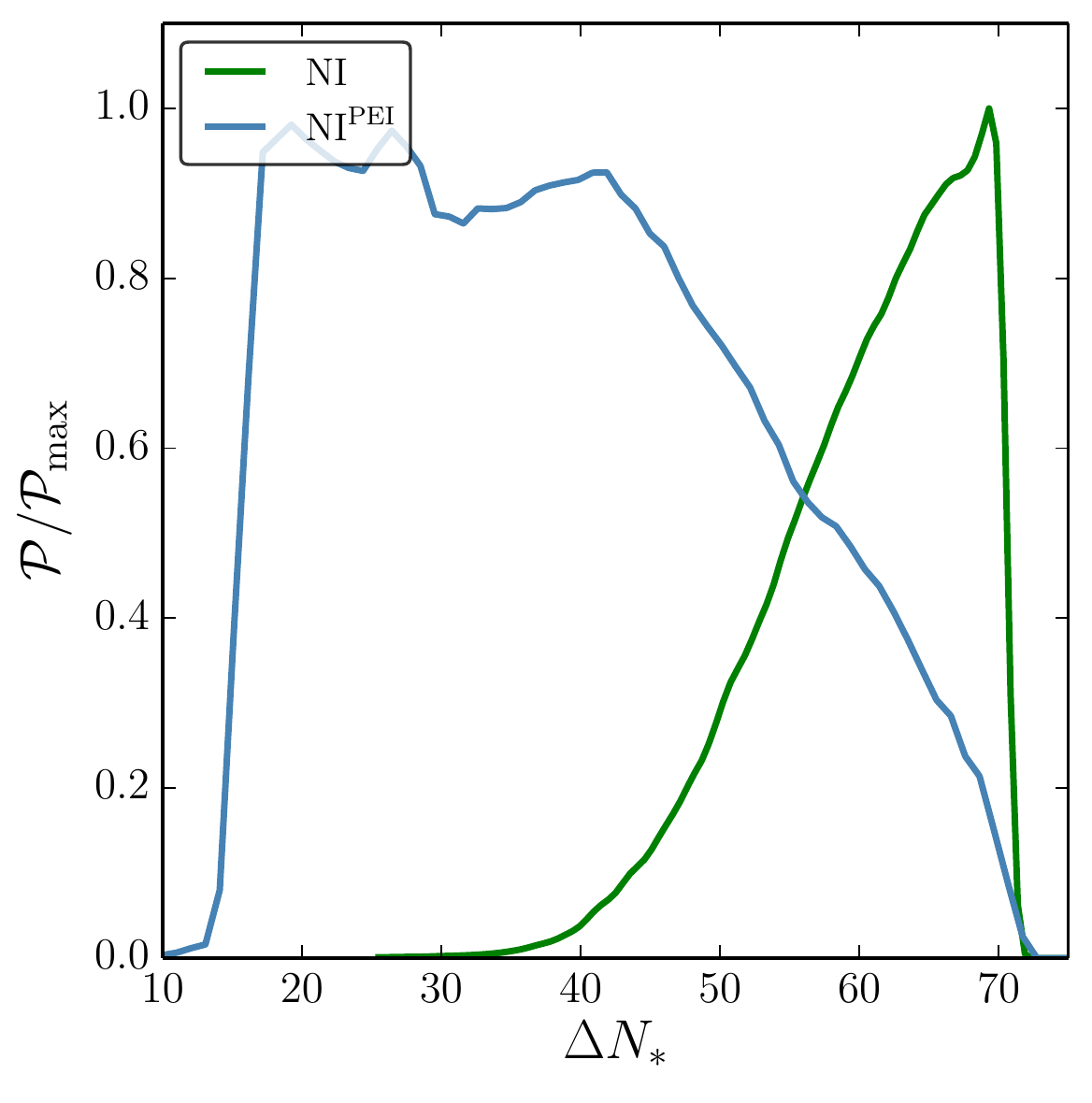}
\includegraphics[width=0.328\textwidth]{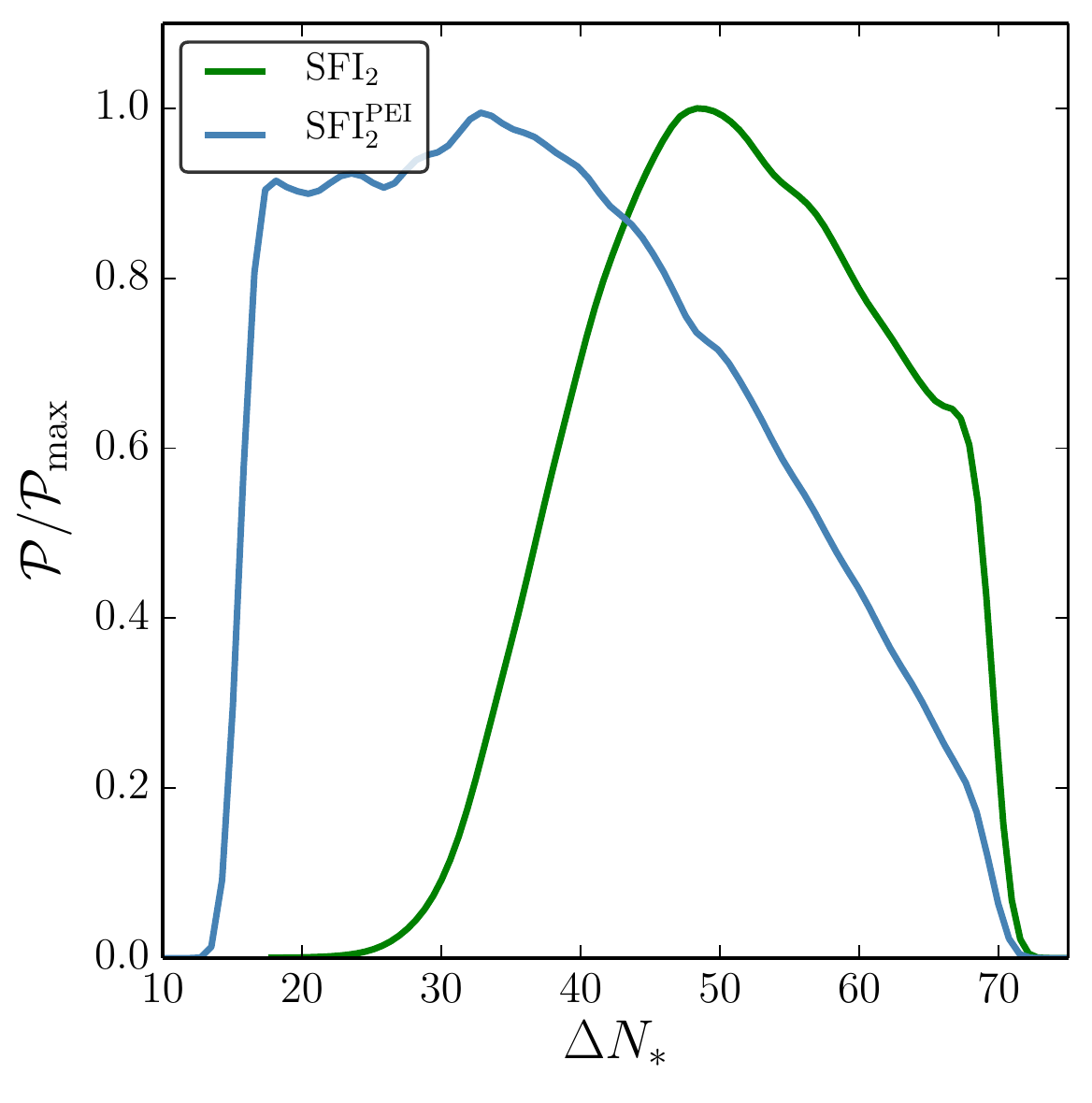}
\includegraphics[width=0.328\textwidth]{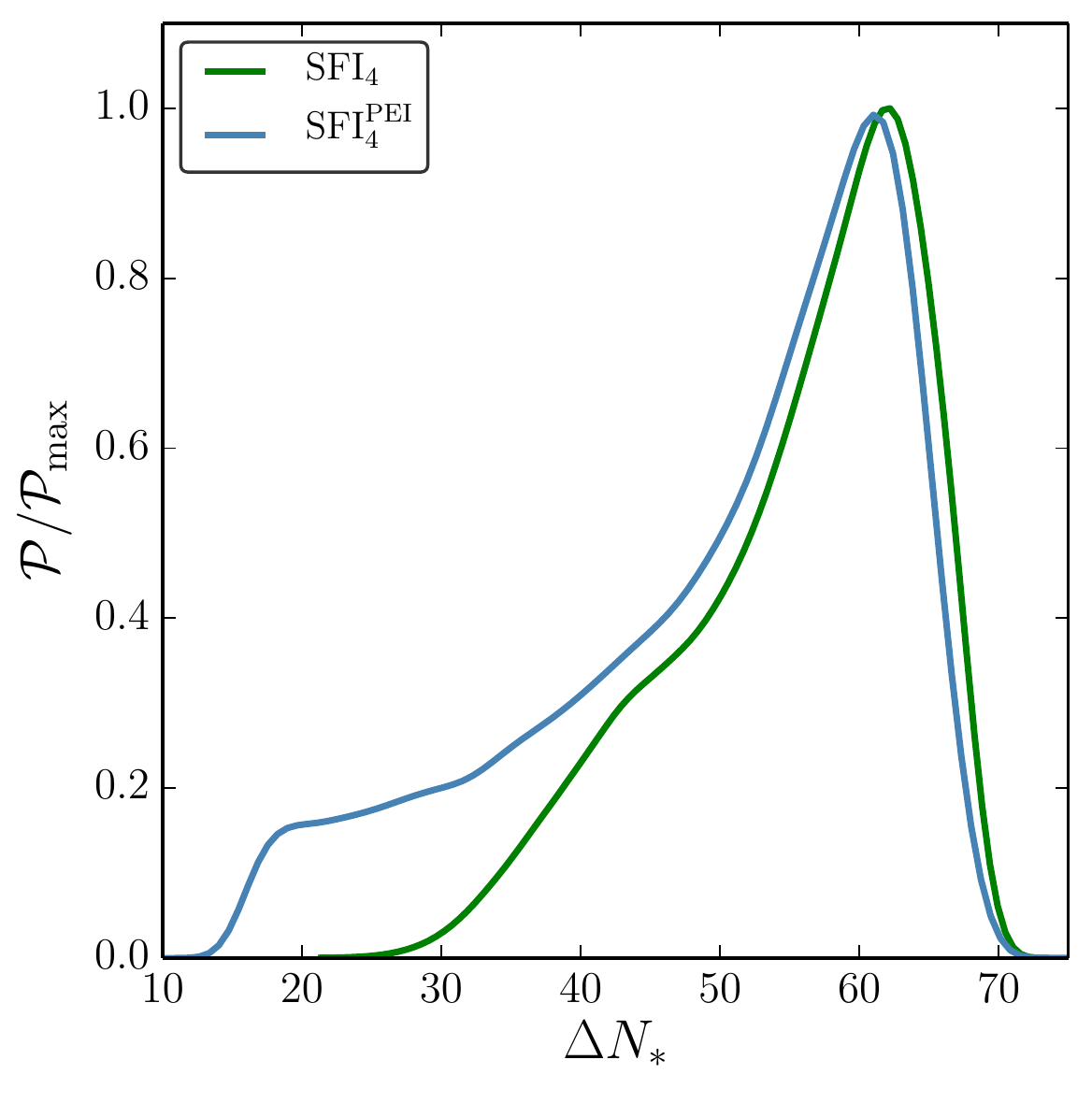}
\includegraphics[width=0.328\textwidth]{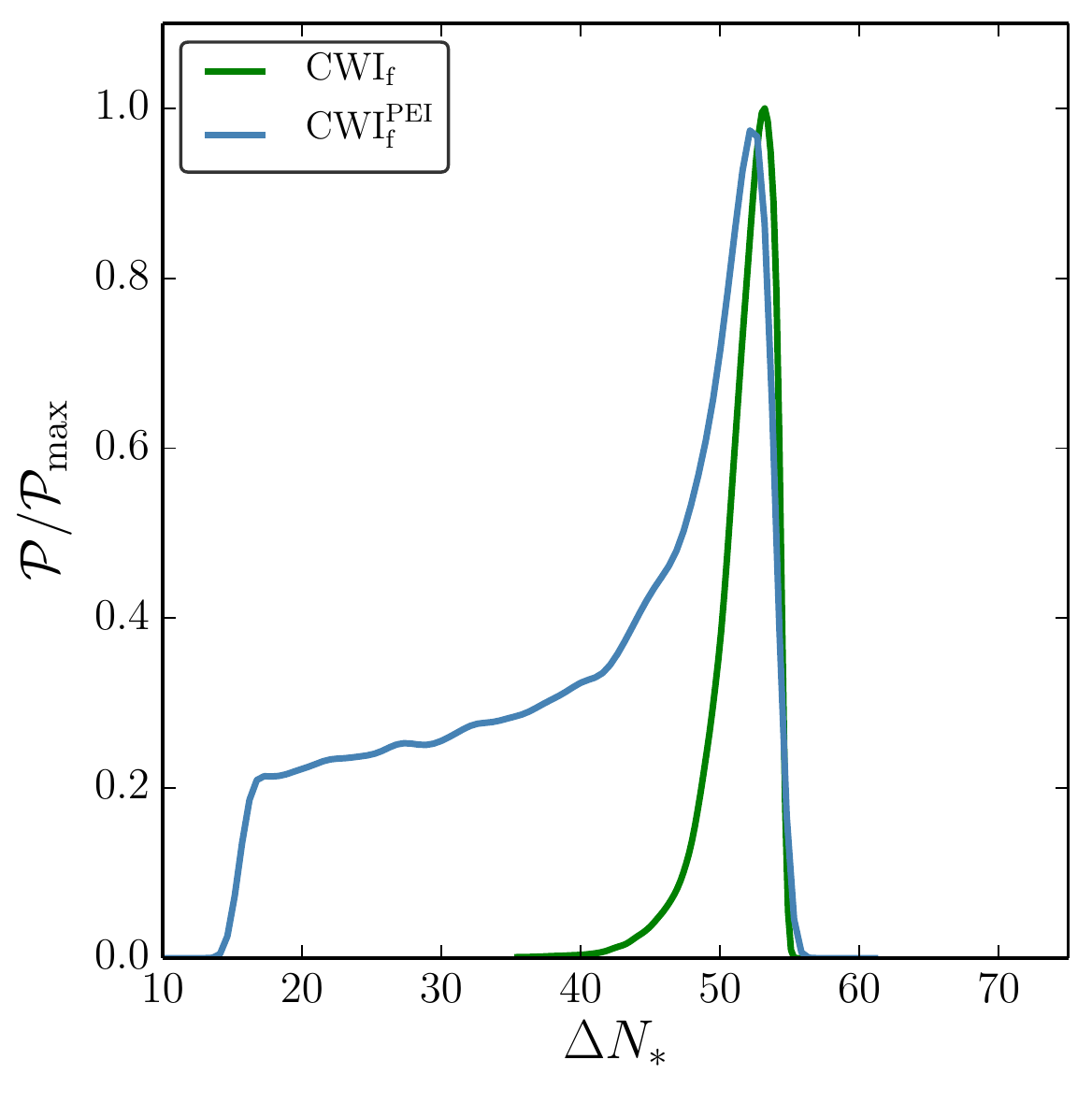}
\includegraphics[width=0.328\textwidth]{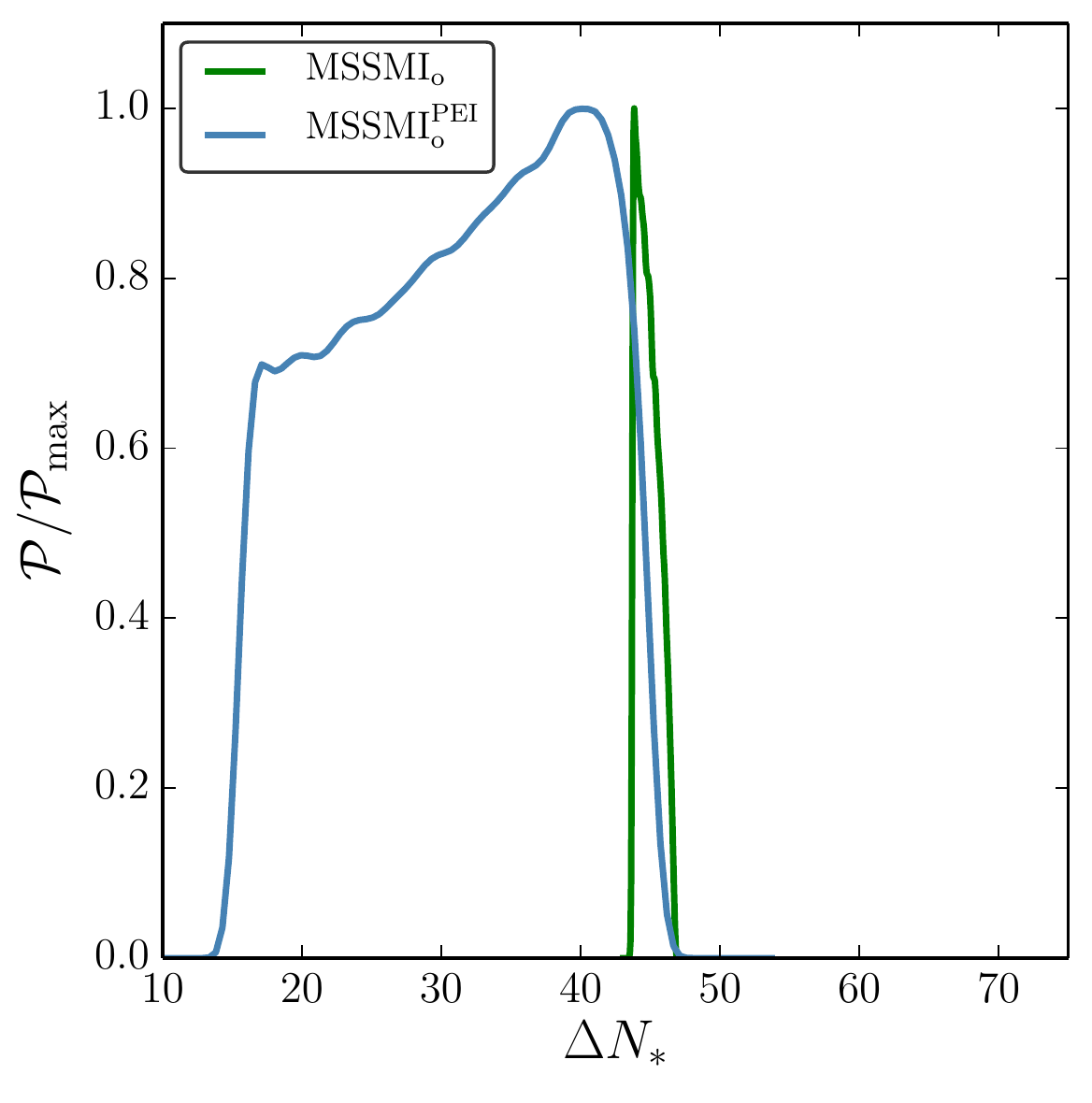}
\includegraphics[width=0.328\textwidth]{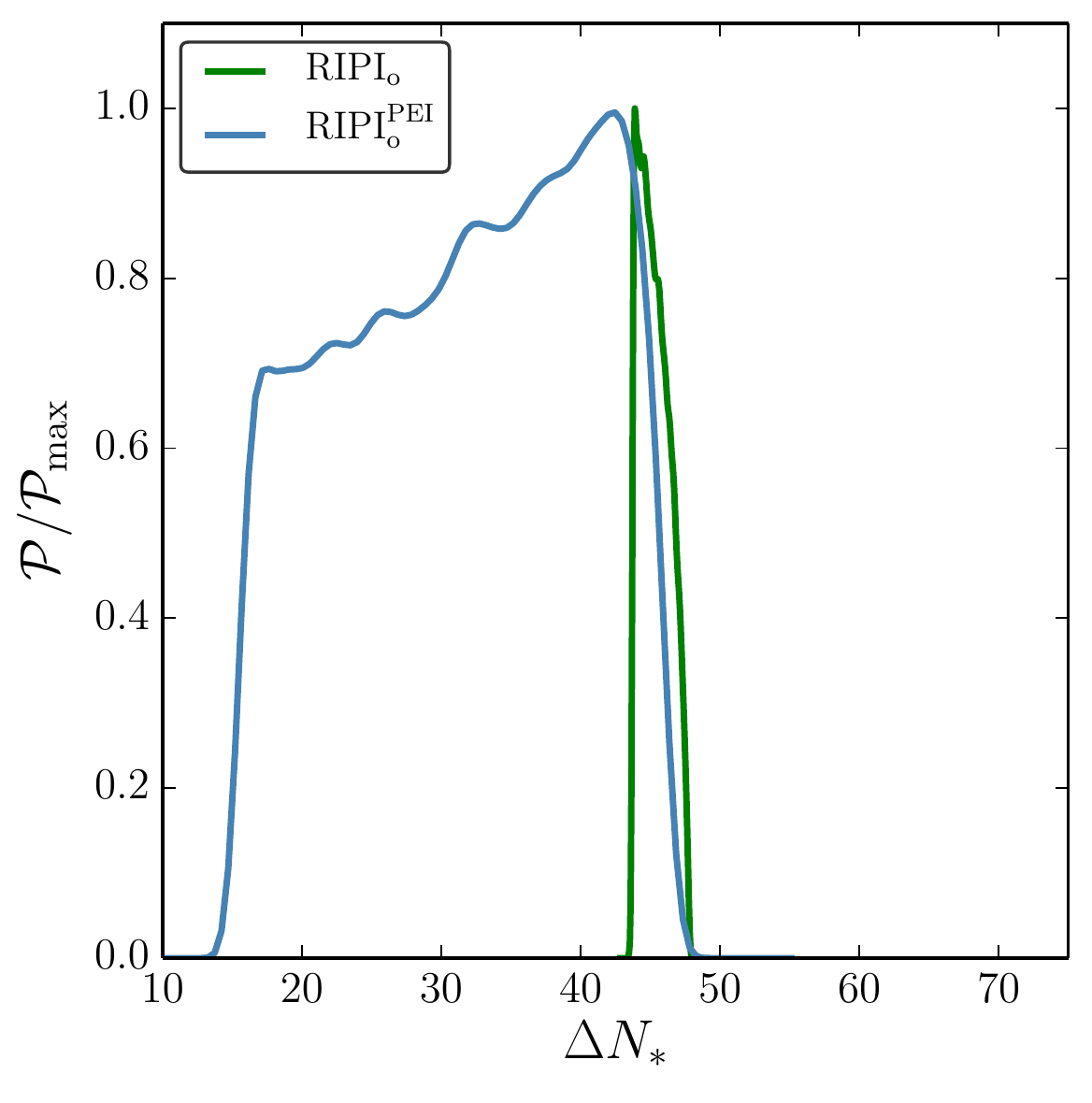}
\caption{Posterior distributions on the number of \efolds~$\Delta N_*$ realized between Hubble exit of the CMB pivot scale $k_* = 0.05\mathrm{Mpc}^{-1}$ and the end of inflation for the models studied in this work, with (blue) and without (green) premature end of inflation (PEI).
\label{fig:efold:posterior}}
\end{center}
\end{figure}

An important aspect of a PEI is that the shift in the observational window it induces is degenerate with uncertainties about reheating, which determines the location of the observational window with respect to the end of inflation. This is why, as explained in \Sec{sec:reheating}, it is important to properly account for the role played by reheating in the analysis. Conversely, introducing a premature end of inflation mechanism also leads to different constraints on the reheating epoch itself. In \Fig{fig:efold:posterior}, we have shown the posterior distributions on the number of \efolds~$\Delta N_*$ elapsed between Hubble exit of the CMB pivot scale and the end of inflation for the models studied in this work, with and without PEI. One can see that in general, PEI allows for a wider range of values of $\Delta N_*$ to be realized, and shows preference for smaller values than the ones obtained in the standard setup where inflation ends by slow-roll violation. This is mainly due to the two following reasons. First, small values of $\Delta N_*$ that are disfavored in the standard setup since they correspond to parts of the inflationary potential too close to the end of inflation where $\epsilon\sim 1$, hence too steep, can be allowed when a PEI is introduced since $\epsilon_\uc$ can be much smaller than one then. Second, in \Eq{eq:DeltaNstar}, one can see that $\Delta N_*$ depends on the absolute energy scale of inflation through $\ln(\rho_*/\Mp^4) = \ln(24 \pi^2 \calP_\zeta \epsilon_*)$ in the second term of the right-hand side, where we have used the expression given above \Eq{eq:lowerbound:MR} for the power spectrum of scalar perturbations $\calP_\zeta$ together with Friedmann equation. Since $\epsilon_*$ is smaller than $\epsilon_\uc$ for the models considered in this work, $\epsilon_*$ is typically much smaller when a premature end of inflation is allowed, which also explains why $\Delta N_*$ is smaller.

The results derived in this work can be interpreted at two different levels. At the first level, the parameterization adopted to describe the GD induces an effective logarithmically flat prior on the first slow-roll parameter at the end of inflation $\epsilon_\uc$ as explained in \Sec{sec:Bayesian}, so that phenomenological and generic constraints about the end of inflation were derived, beyond the mechanism of the GD. They revealed that the Bayesian status of inflationary models can be substantially affected in the presence of a premature termination of inflation, for instance in inflection point models where sharp observational constraints on $\epsilon_\uc$ were derived. At the second level, within the framework of the GD, $\epsilon_\uc$ is related to the mass of the auxiliary field and to the field-space curvature along the inflationary valley. The constraints obtained on $\epsilon_\uc$ can therefore be translated into constraints or measurements on these parameters. Interestingly, we found that these constraints can be quite sharp. For the inflection point models $\mathrm{MSSMI}_\mathrm{o}$ and $\mathrm{RIPI}_\mathrm{o}$ for instance, with $m_h=10 H_\uc$, one obtains $\log_{10}(M_\mathcal{R}/H_*) = 3.0\pm 0.2$. In other words, one can observationally constrain, and even pinpoint, high-energy effects that lie orders of magnitude above the energy scale of inflation, without resorting to primordial non-Gaussianities. This shows how the investigation of ultraviolet effects in the inflationary dynamics, such as the geometrical destabilization where the field-space geometry plays an important role, is crucial to further extend the range of energy scales that are accessible through cosmological surveys.
\section*{Acknowledgements}
S.R-P acknowledges financial support from ``Programme National de Cosmologie et Galaxies" (PNCG) funded by CNRS/INSU-IN2P3-INP, CEA and CNES, France. 
K.T. is partly supported by Grant No.~2014/14/E/ST9/00152 from the National Science Centre, Poland.
V.V. acknowledges funding from the European Union's Horizon 2020 research and innovation programme under the Marie Sk\l odowska-Curie grant agreement N${}^0$ 750491 and financial support from STFC grants ST/K00090X/1 and ST/N000668/1.
\bibliographystyle{JHEP}
\bibliography{GD-Bayesian}
\end{document}